\documentclass{aa}  
\usepackage{graphicx}
\usepackage{txfonts}
\usepackage[colorlinks=true,linkcolor=blue,citecolor=blue,filecolor=blue,urlcolor=blue]{hyperref}
\newcommand{\f}[1]{\fontfamily{qcr}\selectfont{#1}}
\begin{document}
% vajag - pārlasīt uz TENSES & apst v pēc darb v
% epastā ierakstītos labojumus

   \title{A multi-grain multi-layer astrochemical model with variable desorption energy for surface species}
      \author{
		Juris Kalv\=ans	%} \inst{1}
    \and
		Aija Kalni\c{n}a
	  \and
		Kristaps Veitners
		}
   \institute{
Engineering Research Institute `Ventspils International Radio Astronomy Center' of Ventspils University of Applied Sciences,\\
In$\rm \check{z}$enieru 101, Ventspils, LV-3601, Latvia\\
\email{juris.kalvans@venta.lv}
			}
   \date{Received March 20, 2024; accepted May 19, 2024}
  \abstract
% context heading (optional)
{Interstellar surface chemistry is a complex process that occurs in icy layers that have accumulated onto grains of different sizes. The efficiency of the surface processes often depends on the immediate environment of the adsorbed molecules.}
% aims heading (mandatory)
{We investigated how gas-grain chemistry changes when the surface molecule binding energy is modified, depending on the properties of the surface.}
% methods heading (mandatory)
{In a gas-grain astrochemical model, molecular binding energy gradually changes for three different environments -- (1) the bare grain surface, (2) polar water-dominated ices, and (3) weakly polar carbon monoxide-dominated ices. In addition to diffusion, evaporation, and chemical desorption, photodesorption was also made binding energy-dependent, in line with experimental results. These phenomena occur in a collapsing prestellar core model that considers five grain sizes with ices arranged into four layers.}
% results heading (mandatory)
{Variable desorption energy moderately affects gas-grain chemistry. Bare-grain effects slow down ice accumulation, while easier diffusion of molecules on weakly polar ices promotes the production of carbon dioxide. Efficient chemical desorption from bare grains significantly delays the appearance of the first ice monolayer.}
% conclusions heading (optional), leave it empty if necessary
{The combination of multiple aspects of grain surface chemistry creates a gas-ice balance that is different from simpler models. The composition of the interstellar ices is regulated by several binding-energy dependent desorption mechanisms. Their actions overlap in time and space, explaining the similar proportions of major ice components (water and carbon oxides) observed in all directions.}
% up to 6 key words
\keywords{astrochemistry -- molecular processes -- methods: numerical -- interstellar medium: clouds, dust -- stars: formation}
   \maketitle

\section{Introduction}
\label{intrd}

During the last decade, theoretical astrochemists have expanded gas-grain models with additional phases of solid matter. These include molecules residing either on grains of different sizes or in bulk-ice ice layers. The new phases, dynamical evolution of clouds, and detailed microscopic phenomena allow one to paint a new picture about chemical processes in interstellar and circumstellar ices. The microscopic phenomena most notably include efficient chemical desorption (ejection of products from a grain to the gas phase after an exothermic surface reaction) and the molecular desorption (binding) energy that varies depending on its surrounding environment. The above ingredients have never been combined in a single model, which means current astrochemical models may be missing key processes that regulate ice formation and distribution between solid phases.

`Multi-grain' models consider surface chemistry on grains with an assortment of sizes. \citet{Acharyya11} found that the smallest grains, having the largest overall surface, accumulate most of the ice \citep[by default, here we consider the MRN grain size distribution of][used in most studies]{Mathis77}. Ice accumulation onto small grains is amplified by an expansion of the surface area with grain growth, which is most pronounced for the small grains. \citet{Pauly16}, \citet{Ge16}, \citet{Iqbal18}, and \citet{Chen18} consider basic aspects for multi-grain models, such as the effect of the number of grain size bins, the applicability of the rate-equation method, effects of differential grain temperature, and ice accumulation. Several papers focus on the efficiency of cosmic-ray induced desorption for grains of different sizes \citep{Zhao18,Sipila20,KK22}. Some of these indicate that ices cannot efficiently accumulate on the smallest grains \citep{Silsbee21,Rawlings22}. Some of the multi-grain models have been applied in further astrochemical studies \citep{Pauly18,Gavino21}. The chemical effect for the multi-grain approach is that surface processes occur slightly differently for each grain size, owing to their physical differences, such as the temperature and the number of surface adsorption sites.

Compared to multi-grain models, `multi-layer' models have undergone significant evolution. They consider at least two layers of the icy mantle that cover grain surfaces \citep{Hasegawa93m}, both of which may be chemically active \citep{KS10}. Current models resolve separate monolayers \citep[MLs;][]{Taquet12} or up to six chemically active ice layers \citep{Furuya17} with limited-diffusion approaches for binary bulk-ice reactions \citep{Chang14}. The mere existence of subsurface ice is a significant development because it is isolated from most of the desorption mechanisms acting on the exposed surface molecules. Chemically active bulk ice allows for different molecule synthesis paths, for example, in CO- and H$_2$O-dominated environments \citep{Chang16}. Moreover, multi-layer models allow one to regulate evaporation from ices in protostellar envelopes, either via layer-by-layer removal \citep{Taquet14} or by allowing for hyper-volatile molecules to diffuse out of the mantle first \citep{Garrod13}. With hyper-volatiles here we understand icy species with desorption energies $E_D$ below about 1300\,K, such as H$_2$, N$_2$, O$_2$, CO, and CH$_4$ (energies of molecular-level phenomena are expressed in terms of $E/k_B$ per single molecule or atom).

Astrophysical importance for `variable molecular desorption energy' $E_D$ for icy surfaces with a polar (primarily, H$_2$O) and weakly polar (CO and other hyper-volatiles) composition was noted early by \citet{Tielens82}, \citet{Leger83}, \citet{Sandford88}, and \citet{Bergin95}. The latter authors included this effect in an astrochemical model, albeit not in a self-consistent manner \citep{Bergin97}. Compared to hyper-volatile ices, a surface covered with H$_2$O allows for binding via dipole-dipole and dipole-induced dipole interactions, as well as the strong hydrogen bonds. Besides desorption, such bonding affects also mobility of the surface species, and thus, their reactivity. Further exploration of the idea of variable $E_D$ has been limited. The main effects for grains covered by hyper-volatile ices are easier desorption ices and a faster diffusion of molecules and atoms, facilitating chemical reactions \citep{He16,Garrod22}.

Molecule adhesion to the `refractory bare grains' differs from that of ices. An aspect of the variable-$E_D$ approach is differentiating between the bare and ice-covered surfaces. The binding energies to materials similar to interstellar grains are known for a limited number of species \citep[e.g.][]{Vidali91}. Dual (bare grain and ice) $E_D$ have been used by \citet{Chang07}, \citet{Taquet14}, and \citet{Hocuk15}. Unlike multi-layer and multi-grain models, all variable $E_D$ chemical effects are much less understood with no dedicated studies.

The aim of this study is to combine the above phenomena within a single model that produces reasonable results, namely, ice composition, and deduce if variable $E_D$ has astrochemically significant effects. The necessary tasks include
\begin{itemize}
	\item developing an integrated multi-grain multi-layer array system for chemical species, grain, and ice parameters;
	\item adapting or creating descriptions of the microscopic processes, notably, variable $E_D$; the descriptions should allow for simple inclusion in other astrochemical codes;
	\item investigating the significance of variable $E_D$ in modelling results;
	\item exploring effects that arise from phenomena that have not yet been combined together in astrochemical models.
\end{itemize}
Reproducing the proportions of major species observed in interstellar ices has been possible with simpler codes \citep{Ruffle01,Garrod11}, which may have deterred the need for more complex models. The latter two tasks involve limited parameter space analysis and will allow us to understand the gas-grain physico-chemical interplay in dense cloud cores with the updated code. This analysis is essential before further exploration of ice chemistry can occur with the new model.

\section{Methodology}
\label{mthd}

% Table 1
\begin{table*}
\caption{Symbols and abbreviations used repeatedly.}
\label{tab-abbr}
\centering
\begin{tabular}{l l l}
\hline\hline
Abbreviation & Description & Example unit \\
\hline 
\multicolumn{3}{c}{Physical parameters in the gas parcel} \\
\hline 
$t$ & Simulation integration time. & yr, s \\
$T_{\rm gas}$ & Gas temperature. & K \\
$N_H$ & Hydrogen atom column density. & cm$^{-2}$ \\
$n_H$ & Hydrogen atom number density. & cm$^{-3}$ \\
$n(\rm H_2)$ & Hydrogen molecule number density. & cm$^{-3}$ \\
$\zeta$ & Cosmic-ray ionisation rate. & s$^{-1}$ \\
\hline 
\multicolumn{3}{c}{Granular parameters} \\
\hline 
$n_g$ & Number density of interstellar grains. & cm$^{-3}$ \\
$T_d$ & Dust grain temperature. & K \\
$a$ & Dust grain radius. & $\rm \mu$m \\
$b$ & Ice mantle thickness on grains. & $\rm \mu$m, MLs \\
X:H$_2$O & Icy species' X abundance relative to that of water ice. & \% \\
ML & Monolayer -- ice layer with the thickness of a single molecule &  \\
\hline 
\multicolumn{3}{c}{Desorption energy parameters} \\
\hline 
$E_D$ & Surface molecule desorption (binding) energy. & K \\
$E_{\rm diff}$ & Surface molecule diffusion energy; 0.5$E_D$. & K \\
$E_{D,{\rm pol}}$ & Default base $E_D$ value, on H$_2$O ice surface. & K \\
$E_{D,{\rm np}}$ & $E_D$ on hyper-volatile (CO) ice surface. & K \\
$E_{D,\rm bare}$ & $E_D$ for molecules on completely bare grains. & K \\
$E_{D,\rm np}/E_{D,\rm pol}$ & Factor for $E_{D,{\rm pol}}$ change in CO ices. &  \\
$E_{D,\rm pol}$-$E_{D,\rm bare}$ & Energy value of a H-bond. & K \\
$E_A$ & Reaction activation energy. & K \\
$X_{\rm des}$ & Fitting parameter for $E_D$ on mixed ice, Eq. (\ref{vred2}). &  \\
H-bond rule & Bare grain surface does not form hydrogen bonds. &  \\
Hyper-volatiles & Icy species with $E_D$ lower than $\approx$1300\,K. &  \\
\hline 
\multicolumn{3}{c}{Efficiency parameters for desorption mechanisms} \\
\hline 
$f_{\rm CRD}$ & Frequency of cosmic-ray induced heating of whole grains. & s$^{-1}$ \\
$Y_{\rm pd}$ & UV photodesorption yield. & molecule photon$^{-1}$ \\
$f_{\rm CD}$ & Probability for surface reaction product desorption. &  \\
\hline
\end{tabular}
%\tablefoottext{a}{ui}
\end{table*}

The model was developed on the basis of the modified rate-equation code with multi-layer ice chemistry \textsc{Alchemic-Venta} of \citet{K21}, the default reference. Some multi-grain aspects have been tested by \citet{KS22}. The chemical model is set in a gas parcel in a low-mass contracting prestellar dark molecular core. Below, we describe the complete model with all functionalities enabled, referenced to as Model {\f full} in the Results Sect.~\ref{rslt}. Table~\ref{tab-abbr} gives a list of variables and abbreviations used throughout the paper.

\subsection{Chemical model}
\label{mchm}

% Table 2
\begin{table}
\caption{Initial chemical abundances relative to the total hydrogen.}
\label{tab-ab}
\centering
\begin{tabular}{c l}
\hline\hline
Species & X/H \\
\hline
H$_2$ & 0.50\\
He & 0.090 \\
C$^+$ & $1.4\times10^{-4}$ \\
N & $7.5\times10^{-5}$ \\
O & $3.2\times10^{-4}$ \\
F & $6.7\times10^{-9}$ \\
Na$^+$ & $2.0\times10^{-9}$ \\
Mg$^+$ & $7.0\times10^{-9}$ \\
Si$^+$ & $8.0\times10^{-9}$ \\
P$^+$ & $3.0\times10^{-9}$ \\
S$^+$ & $8.0\times10^{-8}$ \\
Cl & $4.0\times10^{-9}$ \\
Fe$^+$ & $3.0\times10^{-9}$ \\
\hline
\end{tabular}
\end{table}

Table~\ref{tab-ab} lists the initial chemical abundances, used at the start of the simulation. The cosmic-ray ionisation rate $\zeta$ was calculated following \citet{Padovani09}, with model `High' spectra from \citet{Ivlev15} and depends on hydrogen column density $N_H$. The $\zeta$ value obtained this way is rather high for the $10^{-17}...10^{-16}$ values typically applied in astrochemistry, hence we divide it by $4\pi$ with the justification that our typical low-mass cloud core is located far from the Galactic centre and is shielded by a parent giant molecular complex. In other words, it can be said that spatially one steradian of the core is exposed to full interstellar cosmic-ray intensity. The intensity of cosmic-ray induced photons depends on $\zeta$ and was calculated with Eq.~(2) of \citet{KK19}. The cloud is irradiated by normal interstellar radiation intensity with $G_0$ of $1.7\times10^8$\,s$^{-1}$cm$^{-2}$, attenuated the cloud's matter with $N_H/A_V=2.2\times10^{21}$\,cm$^{-2}$ \citep{Zuo21}. Gas temperature $T_{\rm gas}$ was calculated according to Eq.~(2) of \citet{K21}. Because this equation works only when the interstellar extinction $A_V$ is below or similar to 40\,mag, $T_{\rm gas}$ was coupled to dust temperature at higher extinctions.

Neutral molecules adsorb onto grain surfaces, forming an ice layer. The sticking coefficient was taken to be unity for heavy species and calculated according to \citet{Thi10} for the light species H and H$_2$. The size of a `cubic average' molecule was assumed to be 0.32\,nm. When the ice thickness $b$ exceeds 1\,ML, excess icy molecules are moved to bulk-ice and are sequentially ordered in three subsurface ice layers. All layers are chemically active. Icy species can be destroyed via photodissociation by interstellar and cosmic-ray induced UV photons at a rate that is equal to 0.3 times their gas-phase photodissociation rate \citep{K18mn,Terwisscha22}. The surface diffusion energy $E_{\rm diff}$ was taken to be 0.50$E_D$. Reactions with activation barriers proceed either by hopping across the barrier or via quantum tunnelling that is possible for H and H$_2$ \citep{Hasegawa93cr}. Chemical reaction rate coefficients were adjusted for reaction-diffusion competition \citep{Garrod11}. Bulk-ice molecules react with other molecules in the same layer with an approach that assumes that they are frozen in place with a bulk-ice binding (absorption) energy equal to 3$E_D$ \citep{K15apj1}. Similar methods for bulk-ice chemistry have recently gained traction \citep{Shingle19,Jin20}.

The model considers several desorption mechanisms, with the simplest being evaporation, which is most important for H$_2$. \citet{Pantaleone21} present a credible evidence that the reaction heat of the common H+H reaction on grains may induce desorption of an adjacent hyper-volatile icy molecule. This indirect reactive desorption mechanism was included in our model with the help of Eq.~(16) of \citet{K15apj1} and an efficiency parameter of $\epsilon=0.001$ desorbed molecules per H+H reaction act \citep[][see also \citeauthor{Takahashi00} \citeyear{Takahashi00}]{Duley93,Willacy94}.

For desorption via cosmic-ray induced whole-grain heating \citep{Hasegawa93cr}, a law that assumes similar heating frequencies for grains of different sizes is used by \citet{KS22}. Here we improved this law according to the exhaustive new data from \citet{KK22}. Namely, the cosmic-ray induced heating frequency $f_{\rm CRD}$ is now proportional to the inverse square root of the grain radius $a$, in addition to its dependence to $N_H$,
\begin{equation}
	\label{ch1}
f_{\rm CRD}(54{\rm K}) = \frac{1.93\times10^{-11}\sqrt{0.05/a}}{A_V^{1.35}4\pi}\,,
\end{equation}
where $a$ is expressed in $\rm \mu$m. The cooling time for the grains was taken to be similar to the characteristic sublimation time of CO \citep{Hasegawa93cr} that is 0.002\,s for a temperature of 54\,K in our model. Like $\zeta$, $f_{\rm CRD}$ was also divided by $4\pi$. Photodesorption and desorption of chemical reaction products are described separately in Sects. \ref{phds} and \ref{chds}.

The chemistry in ice layers was explicitly considered so that the model calculates molecular abundances for each ice layer on each grain type. The actual reaction network includes multiple similar lists of molecular processes for each grain type and ice layer. 

\subsection{Reactions network}
\label{rctn}

% Table 3
\begin{table}
\caption{Additions to thereaction network.}
\label{tab-rea}
%\centering
\begin{tabular}{l c c}
\hline\hline
Gas-phase reactions & $k$(10K)\tablefootmark{a}, cm$^3$\,s$^{-1}$ & Ref.\tablefootmark{b} \\
\hline
CH + CH$_3$OH $\rightarrow$ CH$_3$CHO + H & 2.5E-10 & 1,2 \\
C + H$_2$CO $\rightarrow$ CO + CH$_2$ & 6.2E-10 & 3 \\
CH$_3$ + HCO $\rightarrow$ CH$_3$CHO & 5.0E-11 & 3 \\
\hline
Surface reactions & $E_A$, K & Ref. \\
\hline
CO + H $\rightarrow$ HCO & 2500 &  4\tablefootmark{c} \\
H$_2$CO + H $\rightarrow$ HCO + H$_2$ & 415 &  5 \\
CH$_3$O + H $\rightarrow$ H$_2$CO + H$_2$ & 0 &  5 \\
CH$_2$OH + H $\rightarrow$ H$_2$CO + H$_2$ & 0 &  5 \\
\hline
\end{tabular}
\tablefoottext{a}{Rate coefficient at 10\,K.}
\tablefoottext{b}{1 -- NIST (http://kinetics.nist.gov/kinetics/index.jsp), 2 -- \citet{Johnson00}, 3 -- \citet{Vasyunin17}, 4 -- \citet[][OSU reactions network]{Garrod06}, 5 -- \citet{Minissale16mn}.}
\tablefoottext{c}{Reaction not added, only changed its $E_A$.}
\end{table}

We employ UDfA \textsc{Rate12} chemical network \citep{McElroy13} for the gas phase and a reduced OSU database for surface reactions \citep{Garrod08}. Following \citet{Vasyunin17}, we added gas-phase complex organic molecule (COM) reactions to balance the alcohol-aldehyde chemistry (Table~\ref{tab-rea}). The variable-$E_D$ approach results in generally lower molecular binding energies and more rapid diffusion of surface species that may overproduce CO hydrogenation products. To address this, higher, original OSU database activation energy barrier $E_A$ of 2500\,K was returned for the $\rm CO+H$ surface reaction. Additionally, three `unproductive', H$_2$-producing hydrogenation reactions of intermediate CO hydrogenation products were added, all with branching probabilities of 0.5, as suggested by \citet{Minissale16mn}. The latter reactions supplement similar additions to \textsc{Alchemic-Venta} in Tables 4 and 6 of \citet{K15apj2}. Table~\ref{tab-rea} summarises these changes to the network.

To reduce the overall number of species and reactions, phosphorus compounds with two or more C atoms were removed. These species are irrelevant to the overall chemistry, because of the low abundance of P, and their low abundance, even relative to simpler P species. Network reduction ensured a smooth operation of the code and reduced computing time at a little cost to the scientific output, since we do not study the chemistry of phosphorus.

\subsection{Grain physics}
\label{grn}

% Table 4
\begin{table}
\caption{Dust grain radius $a$ and the number density relative to H atoms.}
\label{tab-gr}
\centering
\begin{tabular}{c c}
\hline\hline
$a$ $\rm \mu$m & $n_g/n_H$ \\
\hline
0.037 & 5.46E-12 \\
0.058 & 1.73E-12 \\
0.092 & 5.46E-13 \\
0.146 & 1.73E-13 \\
0.232 & 5.46E-14 \\
\hline
\end{tabular}
\end{table}

We divided the MRN grain size distribution in five bins with logarithmic spacing. It is a compromise that allows for modelling multi-grain surface chemistry in significant detail, while not making the model overly complex. The number of the grain size bins has a limited effect on modelling results \citep{Iqbal18}, while five bins have been used also by other authors \citep{Acharyya11,Pauly16,Sipila20}. Table~\ref{tab-gr} shows the assumed grain sizes and abundances.

Moreover, here we assumed grains that have already undergone processing in a star-forming region, thus the grains have a carbonaceous coating, not unlike interplanetary dust particles \citep{Flynn20}. While such a choice may be physically justified, in our model it has the benefit of bare surface being significantly different from water-dominated ices. This means that the effects of an $E_D$ that differs for bare and ice-covered grains can be more pronounced (Sect.~\ref{edbr}). A second consequence of the processed-grain assumption is that the smallest grains must have been depleted by sticking to larger grains \citep{Silsbee20}; thus we adopted the sizes and relative abundances of grains from \citet{Sipila20}. Exclusion of the small grains reduces the average temperature and reactivity of the surface species that results, for example, in lower abundances of CO$_2$ ice \citep[see also][]{Iqbal18}. Therefore, the grain size distribution is another aspect that regulates the calculated composition of ices, in addition to the existence of active or passive bulk ice, the $E_{\rm diff}/E_D$ ratio, reaction activation barriers, and selective desorption mechanisms. A benefit for our model is that the exclusion of the small grains allows for adequate operation of the modified rate-equation procedure of the ALCHEMIC code \citep{Semenov10}.

For calculating the cosmic-ray-induced whole-grain heating rate (see above), \citet{KK22} considered refractory grains consisting of 40\,\% amorphous carbon and 60\,\% silicates by mass. The resulting grain density was 2.6\,g\,cm$^{-3}$. The grain mass obtained with this density constitutes 0.4\,\% of the gas mass, in contrast to 0.5\,\% for distributions that include smaller grains, such as \citet{Acharyya11}. We did not consider loss of grain mass; instead the small grains are stuck on the large grains, in effect, increasing their abundance. To account for the mass gap, the \citet{Sipila20} grain abundances were multiplied by a factor of 1.25. An additional 0.8\,\% of the cloud mass are in elements heavier than He (`metals') that constitute icy mantles in freeze-out conditions (Table~\ref{tab-ab}). Grains grow as the icy mantles accumulate and increase in thickness. The temperature of the dust grains $T_d$ was calculated with the method given by \citet[][for $a=0.1\,\rm \mu$m grains]{Hocuk17}, and attributed to different grain sizes following \citet{Pauly16}, that is $T_d\propto a^{-1/6}$.

\subsection{Variable desorption energy}
\label{vred}

For a surface molecule, $E_D$ determines the possibility for desorption. At the low temperatures of molecular clouds, even limited alterations in $E_D$ can determine, if a molecule evaporates or stays on the surface in a given environment. In the case of dense cloud cores, where all species can freeze, except for hydrogen and helium, changes in $E_{\rm diff}$ (a function of $E_D$) are more important, since it determines the rate of surface diffusion and thus, chemical reactions. The synthesis of CO$_2$ ice is a prominent example, with a rate determined by the most mobile of its parent species, CO, that can react with O or OH, both practically immobile at 10\,K, to produce CO$_2$. In the process of a dense core gravitationally contracting, becoming darker and colder, CO becomes immobile, too. The longer CO remains mobile, thanks to a lower $E_{\rm diff}$ on CO-dominated ices, the longer CO$_2$ synthesis is possible. The same effect affects the surface synthesis of some COMs \citep{Vasyunin13}. Therefore, variable $E_D$ for surface species affects the abundances of major as well as minor icy species.

Our task here was to describe $E_D$ as a function of the ice composition, that is, how desorption energy for a molecule surrounded by polar species ($E_{D,\rm pol}$) changes when the same molecule is embedded in a non-polar environment ($E_{D,\rm np}$). In interstellar ices, these two environments are water and the hyper-volatile molecules, respectively. Of $E_D$ measurements and calculations made over the years, we were interested in those that correlate $E_{D,\rm pol}$ and $E_{D,\rm np}$ for the same species. Such measurements are possible only for volatile molecules that evaporate first. Probably the most relevant molecule is CO that is sufficiently abundant to make up the hyper-volatile component \citep{Sandford88,Tielens91}. While it is clear that the actual $E_D$ includes a range of values, depending on the properties of individual adsorption sites and the orientation of the molecule \citep{He18,Grassi20}, here we employ single-value $E_D$ for a species in a given icy phase, a standard practice in astrochemistry.

Desorption energy for CO in watery ices is considered to be 1150\,K \citep{Collings04,Noble12,Penteado17} or 1300\,K \citep{Wakelam17,Das18}. In a pure-ice CO matrix, $E_{D,\rm np, CO}$ of CO was measured to be 954\,K by \citet{Shinoda69}, while more recent measurements give values of 855, 858, 866, and 899\,K \citep[respectively]{Oberg05,Acharyya07,Fayolle16,Martin14}. Another molecule for which data are available is molecular nitrogen. In a H$_2$O matrix, $E_{D,\rm pol, N_2}$ is in the range of 810--1400\,K \citep{Wakelam17,Das18,Penteado17}. For non-polar environments, $E_{D,\rm np, N_2}$, measurements for N$_2$ matrices produce values of 779 and 790\,K \citep[respectively]{Fayolle16,Oberg05}. These laboratory and simulation results yield a wide-ranged $E_{D,\rm np}/E_{D,\rm pol}$ ratio of 0.56...0.98 and average value of 0.8. The values of $E_{D,\rm pol}$ and $E_{D,\rm np}$ clearly differ and do not overlap. The $E_{D,\rm np}/E_{D,\rm pol}$ ratio is similar for CO and N$_2$, justifying an unified approach for variable $E_D$.

The procedure in our model for calculating $E_D$ for species in ices was as follows. The original, or default $E_D$ values correspond to water surface, $E_{D,\rm pol}$. These were adopted from the OSU surface network, replaced by those of \citet{Wakelam17}, where possible. An exception was made for the volatile CO, N$_2$, and CH$_4$ molecules, whose $E_{D,\rm pol}$ were adopted from \citet{Penteado17}. The values of $E_{D,\rm pol}$ correspond to a matrix (surface) consisting of a pure water with $E_{D,\rm pol,H_2O}=5600$\,K. First, the model obtains the weighed average $\bar{E}_{D, \rm pol}$ for the whole ice phase in consideration (one of the four layers on one of the five types of the grains), taking into account all icy species in that phase,
\begin{equation}
	\label{vred1}
	\bar{E}_{D, \rm pol} = \sum^1_{166} X_i E_{D,{\rm pol},i}\,,
\end{equation}
where $X_i$ is the fractional abundance of the $i$-th species in the icy phase, while 166 is the number of unique icy species. Then we calculate $E_{D,\rm np}/E_{D,\rm pol}$ with
\begin{equation}
	\label{vred2}
	E_{D,\rm np}/E_{D,\rm pol} = \frac{\bar{E}_{D, \rm pol}+(X_{\rm des}-1)E_{D,\rm pol,H_2O}}{X_{\rm des}E_{D,\rm pol,H_2O}}\,,
\end{equation}
and the final desorption energy for species A, $E_{D \rm,A}$, used for calculating desorption and surface diffusion rates is
\begin{equation}
	\label{vred3}
	E_{D \rm,A} = E_{D \rm,pol,A} \times E_{D,\rm np}/E_{D,\rm pol} \,.
\end{equation}
There certainly are other possible fitting functions for adjusting species' $E_D$ in accordance with the environment it resides in. Eq.~(\ref{vred2}) depends only on a single fitting parameter $X_{\rm des}$, and allows for mimicking the $E_D$ changes for CO and N$_2$. The value of $X_{\rm des}$ was taken to be 4.00, corresponding to $E_{D,\rm np}/E_{D,\rm pol}\approx0.8$ in accordance with the experimental measurements. For CO, $E_{D,\rm CO}$ decreases from 1150\,K in a H$_2$O matrix environment to 922\,K in a pure CO matrix, while for N$_2$ $E_{D,\rm N_2}$ the respective values are 990 and 786\,K. In modelled multi-layer ices, where Eq.~(\ref{vred2}) operates, the extreme values are never reached and the $E_{D,\rm np}/E_{D,\rm pol}$ lies in between 0.8 and 1. Eq.~(\ref{vred2}) was not applied in cases when a molecule's $E_{D \rm,pol,A}$ was already lower than the average $E_{D \rm,pol}$ in the ice layer. To avoid discontinuities in modelled abundances, variable $E_D$ was applied proportionally to ice thickness, from 0\,MLs with no effect, when the icy molecules barely interact, to full effect at 2\,MLs and above, when most of the icy molecules primarily interact with neighbouring adsorbed species, instead of the refractory grain surface.

Summarising, Eqs. (\ref{vred1}--\ref{vred3}) present a simple approach for calculating $E_D$ for a molecule in ice with changing composition and, thus, changing average desorption energy of species in this icy phase. In practice, $E_{D,\rm np}/E_{D,\rm pol}$ is determined by a few species that dominate a given ice phase at a given time step. Most often these are H$_2$O, CO, CO$_2$, N$_2$, and CH$_3$OH. Eq.~\ref{vred3} was used to calculate $E_D$ for all icy species. Because the composition is different for each icy phase, $E_{D,\rm np}/E_{D,\rm pol}$ was also different for each layer in each grain size bin.

Variable ice $E_D$, calculated from the abundance of H$_2$, has also been employed by \citet{Garrod22}. They do not apparently base their approach on experimental or theoretical data. Within our model, H$_2$ is among the species considered in $\bar{E}_{D, \rm pol}$ and the abundance of the surface H$_2$ itself is regulated with the help of the encounter desorption \citep{Hincelin15}.

\subsection{Desorption energy on bare grains}
\label{edbr}

% Table 5
\begin{table}
\caption{Derivation of the H-bond rule: selected molecular $E_D$ on carbonaceous and icy surfaces.\tablefootmark{a}}
\label{tab-Ed}
\centering
\begin{tabular}{l l l c r}
\hline\hline
Molecule & $E_{D,\rm pol}$, K & $E_{D,\rm bare}$, K & Ref.\tablefootmark{b} & Difference, K \\
 &  &  &  & $E_{D,\rm pol}$-$E_{D,\rm bare}$ \\
\hline
H & 650 & 658 & 1 & -8 \\
H$_2$ & 440 & 542 & 1 & -102 \\
O & 1400 & 1500 & 2 & -100 \\
OH & 3500 & 1360 & 1 & 2140 \\
H$_2$O & 5640 & 2000 & 1 & 3640 \\
O$_2$ & 1000 & 1440 & 1 & -440 \\
O$_2$H & 4300 & 2160 & 1 & 2140 \\
H$_2$O$_2$ & 4950 & 2240 & 1,3 & 2710 \\
CO & 1300 & 1100 & 4 & 200 \\
CH$_3$OH & 3100 & 1100 & 4 & 2000 \\
\hline
\end{tabular}
\tablefoottext{a}{These values were not used in the model. They illustrate the reasoning for estimating the $E_{D,\rm pol}$-$E_{D,\rm bare}$ provided in Table~\ref{tab-Hb}.}
\tablefoottext{b}{1 -- \citet{Cuppen07}, 2 -- \citet{Minissale16aa}, 3 -- \citet{Cazaux10}, 4 -- \citet{Hocuk15}.}
\end{table}
%

% Table 6
\begin{table}
\caption{H-bond rule: assumed $E_{D,\rm pol}$-$E_{D,\rm bare}$ for surface species in the model.\tablefootmark{a}}
\label{tab-Hb}
\centering
\begin{tabular}{l c}
\hline\hline
Molecules or functional groups & $E_{D,\rm pol}$-$E_{D,\rm bare}$, K \\
\hline
HF, NH$_3$, H$_2$O, and H$_2$O$_2$ & 3000 \\
HCl, -OH & 2000 \\
-NH, HCN, and related & 1500 \\
other & 0 \\
\hline
\end{tabular}
\tablefoottext{a}{It is the additional binding energy introduced by H bonds of molecules on ices, compared to molecules on bare carbonaceous grains, see text.}
\end{table}

As stated in Sect.~\ref{grn}, our bare grains have a carbonaceous coating. A few other studies have struggled to obtain an assortment of $E_{D,\rm bare}$ for molecules on carbon because only limited experimental data are available. \citet{Cuppen07} developed such a list for water chemistry that has been updated by \citet{Cazaux10}. Similar lists were compiled by \citet{Hocuk15} and \citet{Minissale16aa}; however the latter studies less rigidly stuck to carbon and included $E_D$ values from experiments with silicate or other materials, instead of using estimates for carbon.

The above-mentioned authors employed small reaction networks. Our task here was to derive a generalised approach with which the differences in $E_{D,\rm pol}$ on ices and $E_{D,\rm bare}$ on carbonaceous grains can be attributed to a variety of species. Table~\ref{tab-Ed} summarises data from studies that have $E_{D,\rm pol}$ and $E_{D,\rm bare}$ based on similar considerations. A striking feature, apparent in the compiled data, is the low $E_{D,\rm bare}$ values for species that are able to form hydrogen bonds. Experiments using highly oriented pyrolytic graphite (HOPG) surface show also a lower bare surface $E_D$ for isolated H-bond forming molecules \citep{Minissale14,Doronin15,Chaabouni18}. Another issue is variation of the desorption energies for volatile species from H to O$_2$, with no clear pattern.

Table~\ref{tab-Ed} allowed us to derive the general approach on devising $E_{D,\rm bare}$: subtracting hydrogen bond energy from $E_{D,\rm pol}$. It gives also an indication on the values that need to be subtracted, which are about $\geq$2000\,K for molecules containing O-H bonds and $\approx$3000\,K for molecules containing two O-H bonds. Naturally, these are lower than the often used hydrogen bond energy of about 2800\,K because, in absence of the H-bonds, other types of molecule-surface bonding are formed instead.

For consistency, we extrapolated the lack of the hydrogen bonds on bare grains, or `the H-bond rule' for other molecules containing electronegative atoms associated with H capable of hydrogen bonding, such as nitrogen or halogens. Such a need is underlined by the study of \citet{Kakkenpara24}, who emphasise the importance of the H-bonds for ammonia in circumstellar ices. The qualitative task was now to choose the types of H-bonds that can make a difference in the grain-ice interface. The quantitative task was to evaluate the differences $E_{D,\rm pol}$-$E_{D,\rm bare}$ for the chosen types of H-bonds.

One has to keep in mind that the carbonaceous grain surface likely is irregular, seeded with heteroatoms with undivided electron pairs and even an occasional H atom attached to an electronegative atom, such as C in \textit{sp} hybridisation. In other words, the carbonaceous surface itself sports some components for weak hydrogen bond formation. From this aspect we infer that only the molecules that can serve as both electron and proton donors, and are able to form the strongest H-bonds can make a difference in the transition from bare surface to icy mantles.

The energy of the OH$\cdots$O bond in water has been studied most extensively \citep[e.g.][]{Tomoda83,Ahirwar22} with dimer dissociation energies usually in the range of 2300 to 3300\,K \citep{Walrafen04,Kikuta08,Spanu08,Sterpone08}. These results allow the effective binding energy difference caused by H-bonding $E_{D,\rm pol}$-$E_{D,\rm bare}$ to be estimated with an accuracy of about 500 to 1000\,K. Organic molecules, for example, alcohols and carboxylic acids are also capable of forming hydrogen bonds via the oxygen atom. Their dissociation energy is in the same range \citep{Andersen15}. Taking into account the data from Table~\ref{tab-Ed}, it seems reasonable to assume $E_{D,\rm pol}-E_{D,\rm bare}=2000$\,K for OH$\cdots$O bonds. It is encouraging that a 2000\,K H-bond energy value is also close to the $E_D$ difference for CH$_2$OH and CH$_3$O obtained by \citet{Garrod08} via completely different considerations.

The \citet{Cuppen07} and \citet{Cazaux10} $E_D$ compilations indicate that the ability of a molecule to form double H-bonds for H$_2$O and H$_2$O$_2$ does not translate into an $E_{D,\rm pol}$-$E_{D,\rm bare}$ twice as high. Here we take this factor to be 1.5 so that $E_{D,\rm pol}-E_{D,\rm bare}=3000$\,K for water and hydrogen peroxide.

In the case of ammonia, NH$\cdots$N hydrogen bonds are only about half as strong. However, the OH$\cdots$N bond energy is similar or even higher than that of OH$\cdots$O, with dissociation energies of about 3800\,K \citep{Yeo94,Kikuta08,Ahirwar21}. Their strength decreases if functional groups are attached to the nitrogen atom \citep{Boryskina07,Vallet15}. The case of ammonia is further complicated by its protonation in water, not explicitly considered here. As a first approximation, we assumed that $E_{D,\rm pol}-E_{D,\rm bare}=3000$\,K for ammonia and 1500\,K for other compounds containing the N-H bond.

For hydrogen halogenides HF and HCl, dissociation energies for H-bonds with water are about 4300\,K and 2700\,K by \citet{Alkorta23}. For hydrogen cyanide, the same authors provide an average value of 2400\,K. In line of the above considerations, we assumed $E_{D,\rm pol}$-$E_{D,\rm bare}$ 3000; 2000, and 1500\,K for HF, HCl, and HCN, respectively. The latter value was attributed also to the related HNC, HNO, and a few other similar molecules.

The $E_{D,\rm pol}$-$E_{D,\rm bare}$ values assumed above are only educated guesses. However, they provide a systemic approach for the bare grain $E_{D,\rm bare}$ that leaves room for improvements with new data. Table~\ref{tab-Hb} summarises the hydrogen bond rule applied in this study. The values provided in the table are exact only for completely bare grains, and their effect is reduced proportionally to ice coverage on grains, until the H-bond rule disappears completely. It occurs, when the formal ice thickness reaches 2\,MLs, and mobility or evaporation can no longer be affected by molecular interactions with the surface of the refractory grains. This means, for example, that a 1\,ML coverage, the $E_{D,\rm pol}$-$E_{D,\rm bare}$ values are only half of their values given in Table~\ref{tab-Hb}. Therefore, during the accumulation of the first two ice MLs on a grain with a given size, the H-bond rule for bare grains was gradually replaced by the variable $E_D$ approach designed for icy environments in Sect.~\ref{vred}.

Two MLs as the final threshold for conversion from bare grain effects to the variable $E_D$ in ices was chosen because at 1\,ML, all molecules are still affected by their attachment to the bare grain surface, and the effects of the latter cannot be discarded yet. Moreover, the higher 2\,ML threshold allowed us to account for some clustering of molecules on bare grain surface \citep{Garrod13mc}, where diffusion, reactions, and desorption on patches of the bare surface are still possible, even when the nominal ice thickness exceeds 1\,ML.

\subsection{Photodesorption}
\label{phds}

% Figure 1
\begin{figure}
%\centering
\hspace{-2cm}
%\vspace{-4cm}
\includegraphics{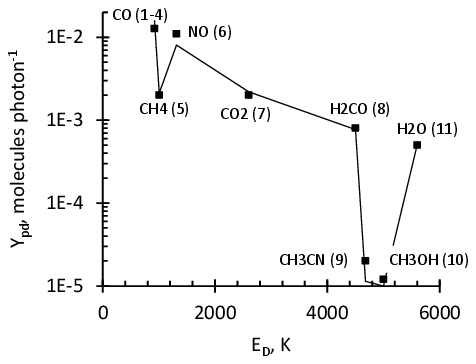}
\vspace{-22cm}
\caption{$E_D$-dependent photodesorption yield of icy molecules. The line connects data points calculated with Eq.~\ref{phds1}, while dots indicate experimental data with references in parentheses: 1 -- \citet{Bertin13}, 2 -- \citet{Fayolle11}, 3 -- \citet{Fayolle13}, 4 -- \citet{Munoz16}, 5 -- \citet{Dupuy17c}, 6 -- \citet{Dupuy17n}, 7 -- \citet{Fillion14}, 8 -- \citet{Feraud19}, 9 -- \citet{Basalgete21}, 10 -- \citet{Bertin16}, 11 -- \citet{Fillion22}.}
\label{fig-pd}
\end{figure}

Photodesorption yield $Y_{\rm pd}$, molecules per UV photon, has now been measured in a number of experiments. Molecular dynamics simulations reveal that it involves photon absorption at different ice layer depths, followed by direct desorption or photodissociation with trapping, recombination, or desorption of products, and `kicking out' neighbours by excited molecules \citep{Andersson06,Arasa15,vanHemert15}. The resulting desorption of intact molecules or their fragments depends on surface type \citep[volatile or non-volatile ices, or bare grain,][]{Bertin12,Potapov19}, composition of icy mixtures \citep{Bertin16,Carrascosa19}, possibility for co-desorption \citep{Bertin13}, as well as the spectrum of the incident radiation \citep{Fayolle11}. Experiments show that $Y_{\rm pd}$ depends on ice temperature \citep{Munoz10,Munoz16}, deposition angle \citep{Gonzalez19}, and ice thickness \citep{Oberg09co,Sie22}. Possible presence of atmospheric gases has to be addressed, while a number of the experiments irradiate their astrophysical ice analogues by photons with energies below 10--11\,eV. This can be a deficiency because important absorption bands may lie at higher energies \citep{Chen14,Martin15,Paardekooper16co}. The missing wavelengths may induce more efficient desorption as in the case of N$_2$, or raise the proportion of dissociative desorption, thus reducing the effective $Y_{\rm pd}$ for intact molecules at full $\lambda$ range.

Several issues become also important, when the experimentally obtained $Y_{\rm pd}$ are applied in astrochemical models. First, two types of UV radiation are present -- interstellar and cosmic-ray induced photons. Some studies differentiate between the two \citep{Fayolle11}; the difference is usually within a factor of 2. Second, dissociated fragments may recombine or react with other surface species and undergo chemical desorption (Sect.~\ref{chds}), enhancing the effective yield. This effect is more pronounced for complex molecules that are more easily dissociated. 

Considering the above, we compiled set of reliable $Y_{\rm pd}$ data, shown in Fig.~\ref{fig-pd}. We used $Y_{\rm pd}$ values for the interstellar radiation field whenever possible because the interstellar photons determine the formation epoch for ices. Moreover, we opted for sources that consider the full 7--13.6\,eV range of UV photons in molecular clouds. Only values for desorption of intact molecules were used because any dissociation fragments have a considerable probability of chemical desorption. Based on the experience from \citet{K15apj1}, temperature and spectral influence were considered to be of minor importance and were ignored. Photodesorption from subsurface layers was addressed by allowing photodesorption for molecule depth from up to 4 ice MLs \citep{Andersson08}.

Carbon monoxide CO photodesorption is, perhaps, the most studied and we had the luxury for obtaining an average desorption yield $Y_{\rm pd}\approx10^{-2}$ value from several experimental reports. A number of papers have studied also the photodesorption of water H$_2$O \citep[e.g.][]{Oberg09h2o,Cruz18,Bulak23}, and carbon dioxide CO$_2$, while only \citet{Fillion14,Fillion22} considered desorption by photons above 11\,eV. Ammonia NH$_3$ was not included because no ice desorption data were found for photons with energies exceeding 10.9\,eV \citep{Martin18}. Molecular oxygen O$_2$ largely desorbs via dissociation, and its yield for intact molecules is uncertain \citep{Fayolle13}. Importantly, usable data are available for some COMs. Their intact molecule $Y_{\rm pd}$ is low and chemical desorption of dissociated fragments is the main ejection pathway \citep{Cruz16,Bulak20}.

An empirical relation that connects the selected measurements is
\begin{equation}
	\label{phds1}
		 Y_{\rm pd} = \frac{7076E_D^{-1.906}}{\eta} \,,
\end{equation}
where $\eta$ is unity for simple molecules with number of atoms $N_{\rm at}<5$ and $\eta=10^{N_{\rm at}-4.2}$ for complex molecules with $N_{\rm at}$ of 5 or more atoms. $E_D$ is expressed in K. This equation is illustrated with Fig.~\ref{fig-pd}. An exception, where Eq.~\ref{phds1} was not applied, was created for the diatomic mono-elemental molecules H$_2$, N$_2$, and O$_2$ that have low yields from pure ices and more efficiently are removed via co-desorption \citep{Fayolle13}. A fixed value $Y_{\rm pd}=0.0055$ was applied for these molecules \citep{Bertin13}. We note that because the molecular $E_D$ varies, photodesorption yields typically are lower by about a factor of 0.8 for hyper-volatile molecules and higher for less-volatile ices. In the important and extreme case of water on bare grains ($E_{D,\rm bare,H_2O}=2600$\,K), its $Y_{\rm pd,bare,H_2O}=0.002$. Such an significantly elevated yield is nonetheless in agreement with experimental data that indicate total (intact and dissociative) $Y_{\rm pd,bare,H_2O}$ of up to 0.5 \citep{Potapov19}.

\subsection{Chemical desorption}
\label{chds}

% Table 7
\begin{table}
\caption{Chemical desorption efficiency as calculated in model for watery ices and bare grains, compared to experimental results.}
\label{tab-cd}
\centering
\begin{tabular}{l | c c | c c l}
\hline\hline
 & \multicolumn{2}{c}{$f_{\rm CD}$ model} & \multicolumn{3}{c}{$f_{\rm CD}$ experiment}  \\
Reaction & ice & bare & ice & bare & Ref.\tablefootmark{a} \\
\hline
$\rm N + N \rightarrow N_2$ & 0.44 & 0.88 & 0.5 & 0.7 & 1 \\
$\rm O + O \rightarrow O_2$ & 0.36 & 0.72 & … & 0.79 & 2 \\
$\rm O + H \rightarrow OH$ & 0.20 & 0.60 & 0.25 & 0.5 & 1 \\
$\rm OH + H \rightarrow H_2O$ & 0.11 & 0.50 & 0.3 & 0.5 & 1 \\
$\rm O_2 + H \rightarrow O_2H$ & 0.00 & 0.05 & … & 0.1 & 3 \\
$\rm CO + H \rightarrow HCO$ & 0.06 & 0.13 & … & 0.1 & 4 \\
$\rm HCO + H \rightarrow CO + H_2$ & 0.18 & 0.36 & … & 0.4 & 4 \\
$\rm H_2CO + H \rightarrow HCO + H_2$ & 0.00 & 0.00 & … & 0.1 & 1 \\
$\rm S + H \rightarrow HS$ & 0.17 & … & $\leq$0.6 & … & 5 \\
$\rm HS + H \rightarrow H_2S$ & 0.10 & … & $\leq$0.6 & … & 5 \\
\hline
\end{tabular}
\tablefoottext{a}{1 -- \citet{Minissale16aa}, 2 -- \citet{Minissale14}, 3 -- \citet{Dulieu13}, 4 -- \citet{Minissale16mn}, 5 -- \citet{Oba18}.}
\end{table}

Highly efficient chemical desorption of exothermic surface reaction products has been explored experimentally during the last decade \citep[][see also \citeauthor{Cazaux10} \citeyear{Cazaux10}]{Chaabouni12}. Desorption probability of this process can be up to 90\,\% (for the $\rm OH + H$ reaction on a silicate surface) and has been quantified and parameterised in further experiments and theoretical works \citep{Dulieu13,Minissale14,Minissale16aa,Fredon17,Oba18,Pantaleone20,Molpeceres23}. This mechanism is especially important for H$_2$O that forms via two-step hydrogenation on grain surfaces. Thanks to hydrogenation-dehydrogenation cycles, chemical desorption is also relevant for CO \citep{Minissale16mn}.

Because chemical desorption is effective for water-forming reactions on bare grains, it significantly affects the onset of the first ice layer formation. This aspect cannot be ignored in chemical modelling focussing on interstellar ices. Thus, we replaced the `reactive desorption' of \citet{Garrod06f}, previously applied in the \textsc{Alchemic-Venta} model, with the `chemical desorption' method by \citet{Minissale16aa}. The variable $E_D$ approach naturally produces different chemical desorption results for different types of surfaces -- bare grains, polar, and hyper-volatile ices. To account for the less-effective desorption from ices observed in experiments, the chemical desorption efficiency (fraction of desorbed molecules) $f_{\rm CD}$ was modified by a factor of 0.5 for reactions on grains with ice thickness exceeding 1\,ML. Table~\ref{tab-cd} compares experimental $f_{\rm CD}$ values to those used in our model. The application of the chemical desorption in concert with variable $E_D$ is what allowed for a realistic and effectual representation of this mechanism.

\subsection{Collapsing prestellar core macrophysical model}
\label{pre}

% Figure 2
\begin{figure}
% \centering
		\vspace{-2cm}
		\hspace{-2cm}
 \includegraphics{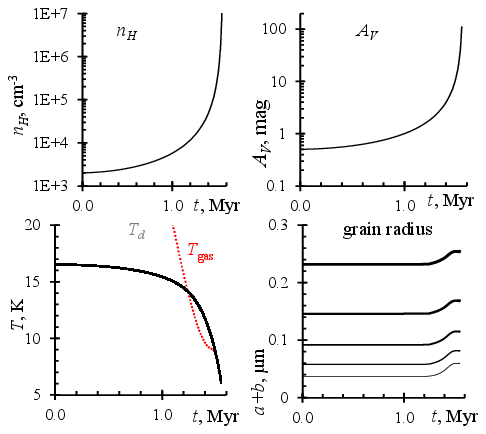}
		\vspace{-20cm}
 \caption{Density, extinction, dust and gas temperature, and grain size evolution in the model. For the latter two graphs, thicker curves are for larger grains.}
 \label{fig-phys}
\end{figure}

We considered a single point located at the centre of a spherical interstellar molecular core. The cloud model is relatively simple and follows the approach of previous studies. It consisted of two main parts. First, the central density $n_0$ of the core was calculated according to a free-fall collapse scenario \citep{Brown88}. Hydrodynamical simulations \citep[e.g.][]{Pavly13,Pavly15} indicate that actual core collapse can be a few times longer than the free-fall time; thus we delayed the contraction rate by a factor of 0.5. The initial conditions were $n_H=2000$\,cm$^{-3}$ and $N_H=1.1\times10^{21}$\,cm$^{-2}$, corresponding to initial interstellar extinction $A_V=0.5$. Second, for each integration step, the spherical (1D) density distribution in the core was obtained with Eq.~(1) of \citet{K21}. This time, core mass was maintained at 2\,M$_\odot$. Core collapse lasts for 1.55\,Myr until a final density of $1\times10^7$\,cm$^{-3}$ is reached. At this point, the freeze-out is effectively over and ice composition does not change any more. Fig.~\ref{fig-phys} shows the evolution of the physical conditions at the centre of the core.

In the first Results subsection~\ref{const} we explored variable $E_D$ effects in pseudo time-dependent Model {\f const} of a stable, dark molecular core, with the collapsing core feature switched off. For chemical relaxation, both models were preceded by a 1\,Myr long diffuse cloud period, with hydrogen numerical density $n_H=2000\,\rm cm^{-3}$ and interstellar extinction $A_V=0.5$\,mag.

\section{Results}
\label{rslt}

% Table 8
\begin{table*}
\caption{Model functionalities and their basic ice chemistry results.}
%\tablefootmark{a}
\label{tab-mod}
\centering
\begin{tabular}{l | c c c c | c c c c c | c}
\hline\hline
 & \multicolumn{4}{c}{Functionality} & \multicolumn{5}{c}{Final ice abundance $n/n(\rm H_2)$} & \\
Model & prestellar & variable &  &  &  &  &  &  &  & $t$ of first \\
 & core & $E_D$ & $f_{\rm CD}$ & $Y_{\rm pd}$ & H$_2$O & CO & CO$_2$ & CH$_3$OH & NH$_3$ & ice ML, kyr \\
\hline
\f const &  & + & $f(E_D)$ & $f(E_D)$ & 1.44E-4 & 3.17E-5 & 5.32E-5 & 1.89E-5 & 1.88E-5 & 46 \\
\f const\_\f{noEd} &  &  & $f(E_D)$ & $f(E_D)$ & 1.43E-4 & 3.10E-5 & 5.37E-5 & 1.85E-5 & 2.00E-5 & 5.2 \\
\f full & + & + & $f(E_D)$ & $f(E_D)$ & 1.28E-4 & 8.08E-5 & 4.37E-5 & 1.26E-5 & 3.41E-6 & 1233 \\
\f noEd & + &  & $f(E_D)$ & $f(E_D)$ & 1.34E-4 & 9.13E-5 & 3.04E-5 & 1.45E-5 & 3.57E-6 & 1176 \\
\f noPD & + & + & $f(E_D)$ & 0.001 & 1.24E-4 & 7.52E-5 & 4.89E-5 & 1.29E-5 & 3.40E-6 & 1192 \\
\f noCD & + & + & 0.03 & $f(E_D)$ & 1.03E-4 & 7.54E-5 & 4.56E-5 & 1.47E-5 & 7.64E-6 & 1034 \\
\hline
\end{tabular}
\end{table*}

Sect.~\ref{mthd} includes a number of assumptions about desorption energy on interstellar grain surfaces. These assumptions may be closer or farther from reality, however, it is clear that molecule diffusion, evaporation, photodesorption, chemical desorption, and desorption by the H+H surface reaction all depend on $E_D$ to some extent. In turn, $E_D$ is subject to change in different surroundings. Our aim was to clarify if this latter dependence is astrochemically significant and deduce its overall character. To do this, we primarily explored results with a model with a complete set of the simulated processes, as described in Sect.~\ref{mthd} (Model {\f full}). For illustrating the significance of one or more processes, limited functionality models were used. Table~\ref{tab-mod} shows that four functionalities of Model {\f full} were switched off or reduced to rudimentary values: the cloud's macrophysical evolution, the variable-$E_D$ approach, chemical desorption, and photodesorption. For context with other models considering multi-grain or bulk-ice chemistry, our Table~\ref{tab-mod} can be compared with Tables 1 and 3 of \citet{Pauly16} and Tables 1 and 7 of \citet{Garrod22}.

We start the description of the results with pseudo-time dependent dense core Models {\f const} and {\f const\_noEd} (Sect.~\ref{const}), continue by describing the results of Model {\f full} and comparing them to Model {\f noEd}, where the $E_D$ variability is disabled (Sect.~\ref{prch}) and conclude by discussing the importance of photo- and chemical desorption (Sect.~\ref{cdpd}).

\subsection{Cold core model}
\label{const}

% Figure 3
\begin{figure}
	%\centering
	\hspace{-2cm}
	%\vspace{-3cm}
\includegraphics{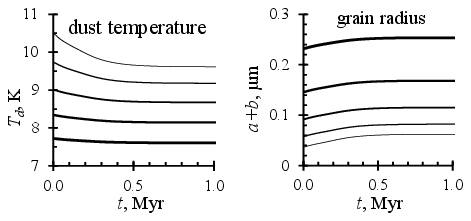}
	\vspace{-23cm}
	\caption{Grain size and temperature in pseudo time-dependent cold core Model {\f const}. Thicker curves are for larger grains; $a$ is the refractory grain radius (constant), while ice mantle thickness $b$ varies with time.}
	\label{fig-const}
\end{figure}

As an initial test case, we present the basic chemical results for a dense, cold, dark core with constant `classical' physical conditions of $n_H=2\times10^4\,\rm cm^{-3}$ and $A_V=10$\,mag. This Model {\f const} was run for an integration time $t=1.0$\,Myr. During the first few hundred kyr, grain growth occurs up to an ice thickness of 78\,MLs on the smallest grains and 66\,MLs on the largest grains. Fig.~\ref{fig-const} shows the change in grain sizes along with the accompanying changes in their temperatures. Gas temperature is constant at $T_{\rm gas}=8.9$\,K, while the cosmic-ray ionisation rate $\zeta=3.2\times10^{-17}$ (Sect.~\ref{mchm}).

% Figure 4
\begin{figure*}
%	\centering
	\hspace{-2cm}
%	\vspace{-1cm}
\includegraphics{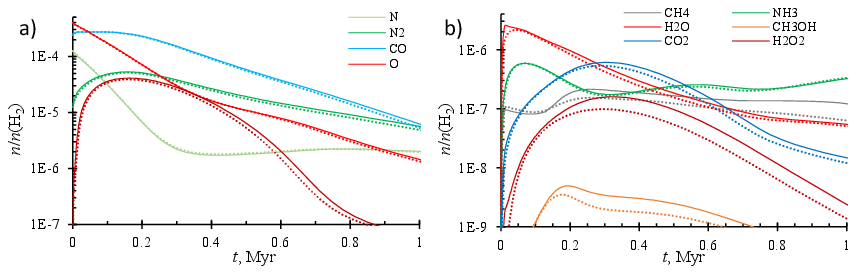}
	\vspace{-22cm}
	\caption{Calculated gas-phase chemical abundances in the cold core model. Panel~(a): relative abundances of major gas-phase species. Panel~(b): gas-phase abundances for important molecules that mostly originate from grain surfaces. In both panels, solid lines are for Model {\f const} with variable $E_D$, while dotted lines are for Model {\f const\_noEd} with an unchanging $E_D$.}
	\label{fig-con-gas}
\end{figure*}

Fig.~\ref{fig-con-gas} shows that the abundances of major gaseous species differ little between Models {\f const} and {\f const\_noEd} with and without the variable-$E_D$ approach, respectively. The easier desorption facilitated by variable $E_D$ is visible in slightly higher gas-phase abundances. The changes are most pronounced for species that are formed via multiple steps on grain surfaces, such as methanol CH$_3$OH and hydrogen peroxide H$_2$O$_2$. Nevertheless, the two simulations produce gas abundances that agree within a factor of 2 for most species and time periods.

% Figure 5
\begin{figure*}
%	\centering
	\hspace{-2cm}
%	\vspace{-3cm}
\includegraphics{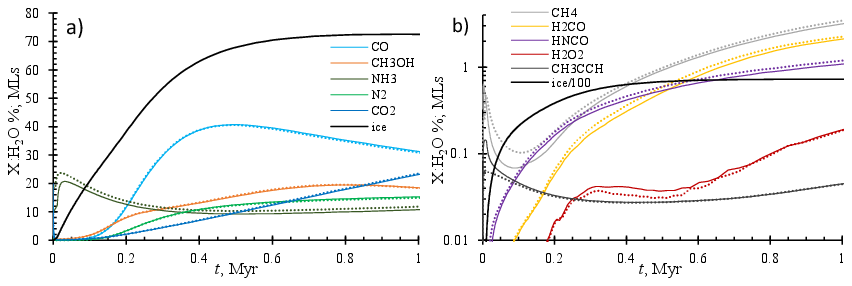}
	\vspace{-22cm}
	\caption{Calculated chemical abundances of icy species in cold core model. Panel~(a): abundances, relative to those of the H$_2$O ice of major icy molecules. The black curve is the average ice thickness \textit{\=b} on grains, expressed in MLs. Panel~(b): other selected abundant ice molecules. For convenience, this time \textit{\=b} was divided by 100. In both panels, solid lines are for Model {\f const} with variable $E_D$, while dotted lines are for Model {\f const\_noEd} without the variable-$E_D$ approach.}
	\label{fig-con-ice}
\end{figure*}

Fig.~\ref{fig-con-ice} shows calculated abundances, relative to water ice, for major icy species for cold core simulations with and without variable $E_D$. In Model {\f const}, 90\,\% freeze-out is reached at 0.45\,Myr that is later probably by 0.1\,Myr or more, relative to comparable multi-grain models \citep{Pauly16,Sipila20}. Differences in the freeze-out times are primarily caused by the consideration or non-consideration of the bulk ice that isolates majority of the icy species from most desorption mechanisms, and by the efficient chemical desorption that acts as delaying function. 99\,\% freeze-out is reached at 0.74\,Myr. For Model {\f const\_Ed} without differing $E_D$ on bare grains and weakly polar ices, the 90\,\% freeze-out is earlier only by about 0.02\,Myr.

A characteristic feature is the low abundance of CO$_2$ ice for most of the time. The ice ratio $\rm CO_2:H_2O\approx20...30$\,\% is often observed to be similar to that of CO:H$_2$O \citep{McClure23} but here reaches only 8\,\% at 90\,\% freeze-out and 16\,\% at 99\,\% freeze-out conditions. However, ice CO$_2$:H$_2$O continues growing to 23\,\% at 1\,Myr thanks to CO$_2$ photoproduction via bulk-ice reactions at the expense of H$_2$O and CO. The cold core model, with its high initial $n_H$ and $A_V$, is more of a testbed calculation, not able to closely represent real-life scenarios. Underproduction of CO$_2$ is a typical feature in such models \citep[see][]{Ruffle01,Bredehoft20}. The similarity between the abundances of the major icy species in Models {\f const} and {\f const\_noEd} indicates that chemistry under rapid freeze-out conditions is regulated mostly by the abundantly adsorbing surface reactants and little affected by $E_D$ variability.

For a review of similar results, albeit without chemical desorption, we refer to \citet{Pauly16}, who also employed a model that considers bulk ice and included a breakdown of ice composition on five grain sizes in their discussion of modelling results. Because we considered larger grains with accordingly lower temperatures, their 5G\_T8 models are most relevant. The main difference between our model and that of \citet{Pauly16} is that we consider chemical processing of the bulk ice that slowly elevates the abundance of CO$_2$ ice. Moreover, our model overproduces methanol CH$_3$OH ice, while their model tends to overproduce methane CH$_4$ ice, which apparently can be caused by differences in reaction networks.

\subsection{General results}
\label{prch}

% Figure 6
\begin{figure*}
%	\centering
	\hspace{-2cm}
%	\vspace{-3cm}
\includegraphics{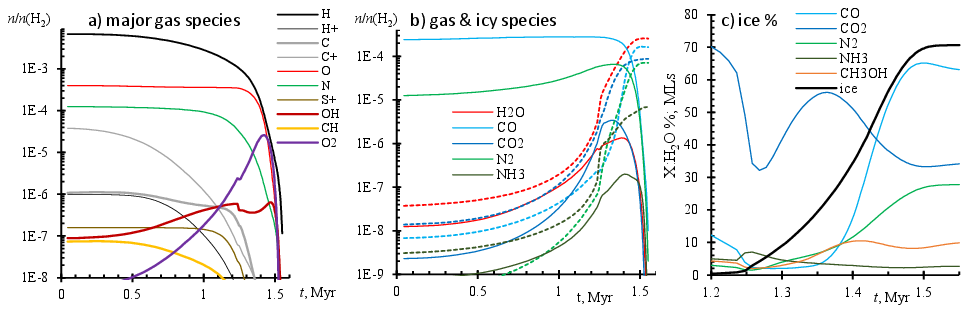}
	\vspace{-23cm}
	\caption{Overall chemical results for prestellar core Model {\f full} with variable $E_D$ and other features enabled. Panel~(a): abundance of the primary gaseous species relative to that of H$_2$. Panel~(b): species that are abundant in ices; solid lines are for the gas phase, while dashed lines are for solid phase abundances, relative to H$_2$. Panel~(c):	percentage, relative to H$_2$O ice, for major icy molecules. The black curve is the average ice thickness \textit{\=b} on grains.}
	\label{fig-gen}
\end{figure*}

As a context for the discussion that follows, in Fig.~\ref{fig-gen} we show the general chemical results -- abundances of major species -- for the prestellar core model. With regard to evolution of ices, three periods can be discerned. First is the translucent cloud, dominated by gaseous atomic species and CO, while ice thickness remains below 1\,ML. This period lasts for $\approx$1.2\,Myr, until $n_H$ exceeds 10$^4$\,cm$^{-3}$ and $A_V\approx1.8$\,mag. Second is the ice formation period, when the core contraction becomes increasingly rapid and up to 99\,\% of the metals are accreted onto grains at $t$=1.5\,Myr, $n_H=3\times10^5$\,cm$^{-3}$ and $A_V=14$\,mag. Third, during the remaining 25\,kyr density increases thirtyfold (with a presumed further collapse towards the first core) with little change in the composition of ices. This period is of limited interest for this study. We continue by describing the translucent and ice formation periods in more detail.

\subsubsection{The sub-ML regime in translucent cloud}
\label{submm}

% Figure 7
\begin{figure*}
%	\centering
%	\hspace{-2cm}
%	\vspace{-2cm}
\sidecaption
\includegraphics[width=11cm]{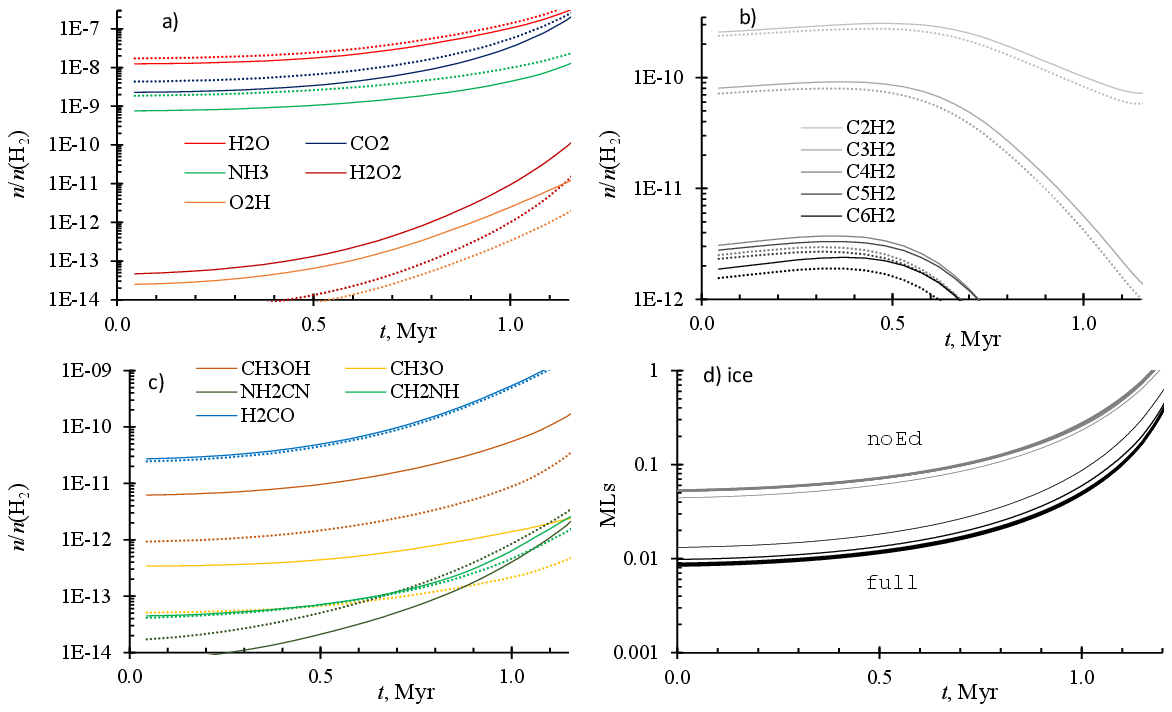}
%	\vspace{-10cm}
	\caption{Variable-$E_D$ induced changes in the translucent cloud. Comparison between Models {\f full} (solid lines) and {\f noEd} (dotted lines) for selected species including those with gas-phase abundances most affected by $E_D$ changes via the H-bond rule: inorganic in panel~(a), carbon chains in panel~(b), and another organics in panel~(c). Panel~(d) shows the growth of the ice mantles on the grains, with thinner lines indicating smaller grains. The changes are caused by a lack of strong H bonding on bare grains in Model {\f full}.}
	\label{fig-trans}
\end{figure*}

Adsorbed molecules with average ice thickness below 1\,ML likely are present in diffuse and translucent molecular envelopes and thus form a part of sight-lines towards prestellar cores. In Model {\f full}, this is the period when the H-bond $E_D$ rule of the bare grains is in effect (Sect.~\ref{edbr}). The lack of the hydrogen bonding for surface species does not lower their $E_D$ and $E_{\rm diff}$ to significantly promote evaporation or diffusion across the grains (Table~\ref{tab-Hb}). However, it is sufficient to elevate chemical desorption and photodesorption yields (Sects. \ref{phds} and \ref{chds}).

Surface chemistry in the translucent cloud is regulated by an interplay between accretion and desorption. Accreted molecules can be desorbed directly by photodesorption or dissociated into fragments. Chemical radicals created on the surface or accreted from the gas react and the product molecules can be desorbed via chemical desorption. Fig.~\ref{fig-trans} shows that two groups of surface-origin species can be discerned, whose gas abundance is regulated by desorption. First, species like H$_2$O, NH$_3$, CH$_4$, CO$_2$, CH$_3$OH reach high gas-phase abundances relative to H$_2$ ($n/n(\rm H_2)$) in excess of $10^{-9}$, thanks to photodesorption and rather high surface abundances of $>10^{-8}$. Second, a variety of species have highly efficient chemical desorption that, in combination with surface photodissociation, increase gas-phase $n/n(\rm H_2)$ to $10^{-12}$ and above. These include, for example, hydrogen peroxide H$_2$O$_2$ that is subject to a strong H-bond rule (Table~\ref{tab-Hb}) and whose abundance changes by about an order of magnitude between Models {\f full} and {\f noEd}.

In absolute numbers, most of the COMs have gas-phase $n/n(\rm H_2)$ below $10^{-13}$ during the translucent period. However, for methanol and a few related compounds, such as CH$_3$O, abundances exceed $10^{-12}$ and thus the effect of the lack of H bonds on bare-grains could be observed. Fig.~\ref{fig-trans} shows that formaldehyde H$_2$CO that has also effective chemical desorption. Its $E_D$ is unaffected by the H-bond rule that means it has similar translucent cloud abundances for Models {\f full} and {\f noEd}. The abundance of CH$_3$OH and CH$_3$O is higher by a factor of $\approx5$ in Model {\f full} relative to Model {\f noEd}; for H$_2$O$_2$ and O$_2$H this factor is 6...10. These factors become lower as time goes by because with the accumulation of ice, the importance of the H-bond rule decreases.

Lack of the H bonds on bare grains has a significant effect on overall ice abundances in the sub-ML regime. The release of more H$_2$O, CO$_2$, and NH$_3$ to the gas phase in Model {\f full} results in lower by a factor of 2...3 adsorbed species' abundances in the variable $E_D$ model. Overall, the H-bond rule delays the formation of the first ice monolayer by 40\,kyr for the smallest and 67\,kyr for the largest grains. This delay has an inverse effect on H$_2$O and CO$_2$, for which the gas-phase abundances are lower by a factor of 2 because there are less of these molecules on the surface, available for photodesorption. In turn and to a similar extent, a lower gas-phase H$_2$O abundance positively affects the abundance of carbon chains because H$_2$O interferes with carbon-chain gas-phase production by reducing the abundance of their building blocks, such as CH, as illustrated with panel~(b) of Fig.~\ref{fig-trans}. The increase in carbon chain abundance by about a factor of two may not be high but its significance lies in that virtually all unsaturated chains are affected by it.

\subsubsection{Ice formation epoch}
\label{epch}

% Figure 8
\begin{figure*}
%	\centering
	\hspace{-2cm}
%	\vspace{-2cm}
\includegraphics{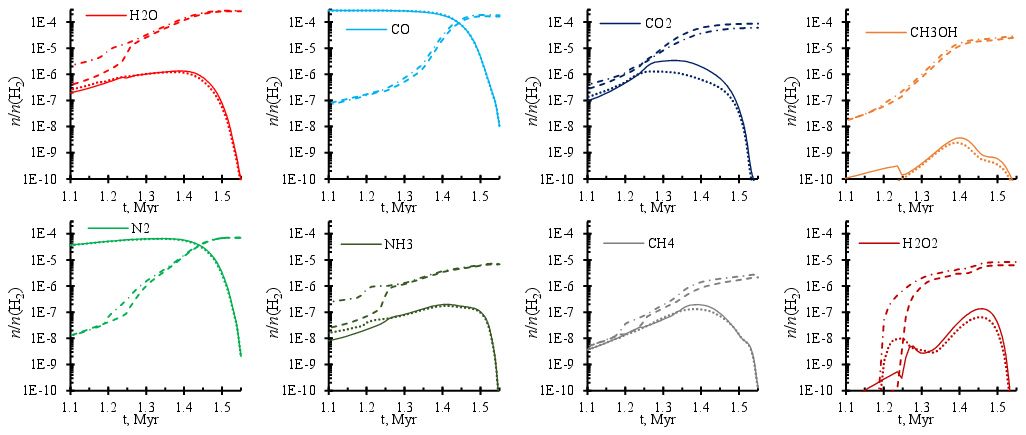}
	\vspace{-19cm}
	\caption{Variable-$E_D$ induced changes during ice formation epoch: calculated $n/n(\rm H_2)$ for species ending up with a high ice abundance. Solid and dashed lines are for the Model {\f full} gas phase and solid species, respectively. Dotted and dash-dotted lines are for gas and solids in Model {\f noEd}, respectively.}
	\label{fig-epoch}
\end{figure*}

The ice formation period is when most of the ice mass is being accreted onto grains and the ice acquires its initial composition. If the ice is not destroyed (e.g. by falling into the protostar), this composition can be further modified by heating or photoprocessing. Ice formation is characterised by initial formation of a 1\,ML thick H$_2$O-CO$_2$ layer at the end of the translucent cloud period. It is followed by further accumulation of H$_2$O and CO$_2$. When $T_d$ drops below 12\,K and CO becomes immobile, CO$_2$ surface synthesis stalls \citep{Pauly16} and CO ice accumulates more rapidly, eventually overtaking CO$_2$ but not H$_2$O.

In Model {\f full}, a lower (typically by about 10\,\%, i.e. $E_{D,\rm np}/E_{D,\rm pol}\approx0.9$) varying $E_D$ has two counteracting effects on the abundance of the major icy species, compared to Model {\f noEd} with unchanging $E_D$. First, a lower $E_{\rm diff}$ for surface CO allows it to remain mobile for longer in Model {\f full} and thus produce more CO$_2$ in reactions with O and OH. This effect becomes visible when $T_d$ drops below 14\,K. Second, lower $E_{D, \rm CO}$ allows for a more efficient desorption of CO, retaining it longer in the gas-phase. As a result, CO accretes later at lower $T_d$ and produces less CO$_2$. The balance of these two effects depends on the evolution of the modelled cloud and the choice of model parameters, such as the efficiency of the various desorption mechanisms, grain size distribution, $T_d$ of grains of different sizes, and, in our model, also the variable $E_D$ approach. Table~\ref{tab-mod} shows that Model {\f full} has a CO$_2$ ice abundance higher by a factor of 1.4 than Model {\f noEd}, that is, CO$_2$ production at lower grain temperatures has been more significant.

The overall effect of the addition of the varying $E_D$ in the model during the freeze-out stage is higher gas-phase abundances for most species, typically elevated by a factor of 1.5...2 (abundance ratio Model {\f full/noEd}). This occurs thanks to the more efficient chemical and photodesorption. Fig.~\ref{fig-epoch} demonstrates several general variable-$E_D$ effects that affect major icy species. 

First, the period between 1.19 and 1.26\,Myr differs most because the first ice MLs have formed in Model {\f noEd} but not on the grains of Model {\f full}. The synthesis of H$_2$O, NH$_3$, CO$_2$, and CH$_3$OH depends on intermediate radical species with  hydrogen bonds (OH, NH, NH$_2$, CH$_2$OH). The non-existence of strong H-bonds on bare surface and the resulting efficient chemical desorption (Table~\ref{tab-cd}) is what delays the accumulation of the ice layer in Model {\f full}.

Second, the lower overall $E_D$ in ices continues to heighten $f_{\rm CD}$ and $Y_{\rm pd}$ for the remainder of cloud evolution, ensuring higher gas-phase abundances for CO, CO$_2$, N$_2$, and CH$_4$ in Model {\f full}. A $E_{D, \rm CH_4}$ lowered by about 100\,K means that desorption by the H+H surface reaction heat works on methane in Model {\f full} but not in Model {\f noEd}.

The third effect is chemistry in bulk-ices, the primary place of synthesis for hydrogen peroxide H$_2$O$_2$ and contributes also to the formation of CO$_2$. Bulk-ice synthesis becomes possible only when $>$1\,MLs of ice have formed. It switches on rapidly and has an immediate effect on H$_2$O$_2$ abundances in ice. H$_2$O$_2$ appears also in the gas because our model allows for photodesorption of bulk ice equal to up to 3\,MLs, in addition to 1 surface ML (see Sect.~\ref{phds}).

\subsubsection{Distribution of ices}
\label{grlay}

% Figure 9
\begin{figure}
%	\centering
	\hspace{-2cm}
%	\vspace{-2cm}
%\includegraphics[width=\columnwidth]{fig-ice-size.eps}
\includegraphics{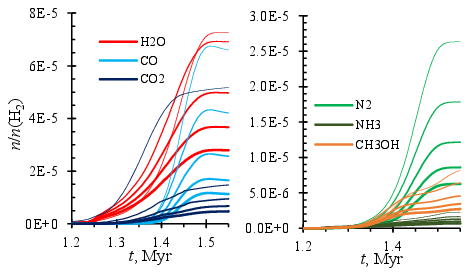}
	\vspace{-22cm}
	\caption{Evolution of the relative abundances for major icy species on grain populations with different sizes. Thicker lines are for larger grains.}
	\label{fig-ice-size}
\end{figure}
%

% Figure 10
\begin{figure}
%	\centering
%	\hspace{1cm}
%	\vspace{-2cm}
\includegraphics[width=16cm]{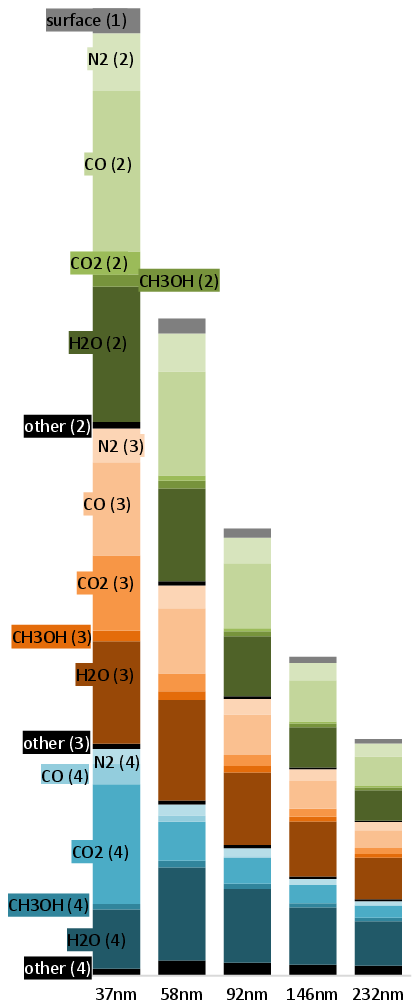}
	\vspace{-7cm}
	\caption{Distribution of major icy species in different grain size bins, indicated by their size (nm) at final time $t=1.55$\,Myr. The four ice layers are numbered: layer (1) is the surface, while layer (4) is adjacent to the refractory grain core. Species within a single layer are intermixed, they are shown separately here to illustrate their proportions within the layer. Although thickness is similar for all bulk-ice layers for a given grain size, the species in the outer layers are more abundant because the grain has grown. This effect is more pronounced for the smallest grains. Most of the `other' molecules are NH$_3$ and also CH$_4$.}
	\label{fig-lay}
\end{figure}

During the $\approx$50\,kyr after the ice formation epoch, residual gas molecules continue to be depleted onto grains. No equilibrium is established because gas density continues to increase rapidly. The final ice composition significantly differs between separate grain size populations. Fig.~\ref{fig-ice-size} shows that all grains achieve a similar ice thickness in the range of 70...75 MLs. Notably, at the end of the simulation with variable $E_D$, the smallest grains carry 59\,\% of all CO$_2$ and 40\,\% of CO ice, while other icy species are more evenly distributed between all grain size bins. These results can be explained primarily by the higher temperature of the small grains that make CO$_2$ surface synthesis faster and possible for a longer time in cloud evolution. Because a single CO$_2$ molecule is formed instead of two molecules of H$_2$O and CO, ice layer on small-grains is not the thickest (72\,MLs at $t=1.55$\,Myr). Concentration of CO$_2$ on smaller grains occurs also in other multi-grain models \citep{Pauly16,Iqbal18}.

Fig.~\ref{fig-lay} illustrates the proportions of the major icy species in the grain size bins at simulation end time $t$=1.55\,Myr. The smallest 0.037\,$\rm \mu$m grains carry the highest amount of ice -- 37\,\% of all molecules, compared to only 9\,\% on the largest 0.232\,$\rm \mu$m grains. Unlike other simulations, especially two-phase models without bulk-ice \citep{Iqbal18,Sipila20}, cosmic-ray induced desorption has no major effects on the distribution of ices between grains of different sizes.

\subsection{Effects of $E_D$-dependent chemical- and photodesorption}
\label{cdpd}

% Figure 11
\begin{figure*}
%	\centering
%	\hspace{-2cm}
%	\vspace{-2cm}
\includegraphics{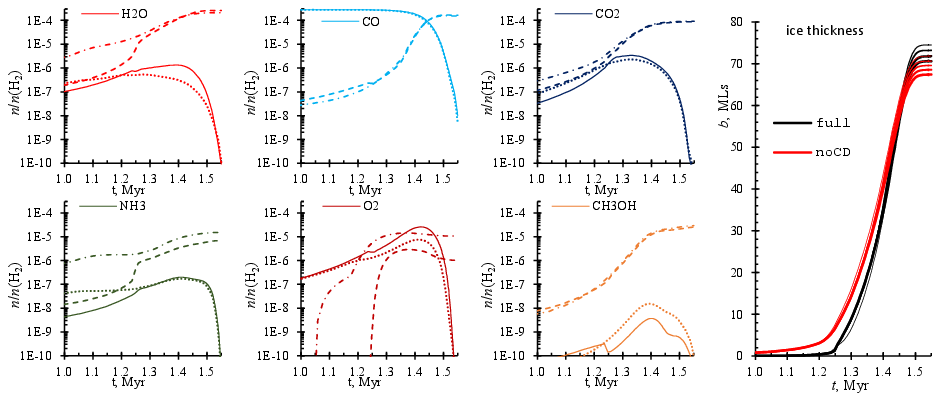}
	\vspace{-20cm}
	\caption{Ice growth and chemical desorption. This is a comparison of abundances between Model {\f full} gas (solid lines) and icy (dashed lines) species with those of Model {\f noCD} (dotted and dash-dotted lines for gas and ices, respectively). In the ice thickness plot, thicker lines are for larger grains, according to Table~\ref{tab-gr}.}
	\label{fig-xcd}
\end{figure*}

Photodesorption and chemical desorption (Sects. \ref{phds} and \ref{chds}) are two mechanisms whose yields quantitatively depend on $E_D$ and are calculated separately for each surface molecule in each of the five grain size bins in the programme. Thus, $f_{\rm CD}$ and $Y_{\rm pd}$ change along with $E_D$, whose variation is described in Sects. \ref{vred} and \ref{edbr}. In other words, chemical desorption and photodesorption are the instruments that help communicate $E_D$ variations to the gas-ice balance. The efficiency of these mechanisms is anchored in experimental data and often is about an order higher than the safe assumptions applied in astrochemical models during preceding decades.

For comparison with Model {\f full}, we ran simulations with the following changes:
\begin{itemize}
	\item Model {\f noPD}, where the $E_D$-dependent photodesorption yield was replaced with a single constant value $Y_{\rm pd}=0.001$ for all species;
	\item Model {\f noCD}, where the $E_D$-dependent chemical desorption efficiency was replaced with constant 3\,\% of all the surface reaction products going to the gas phase \citep[$f_{\rm CD}=0.03$; a simplified version of the reactive desorption by][]{Garrod06f}. This $f_{\rm CD}$ is still significantly higher than that used by \citet{Garrod22}.
\end{itemize}
The above means that in Models {\f noPD} and {\f noCD}, photo- or chemical desorption are not disabled, only significantly reduced for simple icy molecules. The fixed $f_{\rm CD}$ and $Y_{\rm pd}$ values are close to previously commonly used desorption parameters. For complex molecules with high $E_D$ and high number of atoms, these fixed desorption efficiencies are actually higher, when compared to the fully $E_D$-dependent Model {\f full}. This is why Model {\f noCD} shows a peak gas-phase methanol abundance of $1.5\times10^{-8}$ relative to H$_2$, four times higher than that of Model {\f full} (Fig.~\ref{fig-xcd}).

In Model {\f noPD}, the first ice ML is formed at 1.19\,Myr on the smallest grains and 1.23\,Myr on the largest grains, compared to 1.23...1.24\,Myr for Model {\f full}. Such a moderately earlier ice layer formation is associated with build-up of solid CO$_2$ and H$_2$O ices that have translucent stage abundances higher by factors of 3 and 2 relative to Model {\f full}, respectively, thanks to their lower $Y_{\rm pd}$ on bare grains. Because CO$_2$ formation occurs mostly on the smallest grains because of their higher $T_d$, the first ice layer on the small grains in Model {\f noPD} forms 40\,kyr earlier than in Model {\f full}. For other grain size bins this difference is 16\,kyr. Thanks to this advantage, the smallest grains grow the thickest ice, 77\,MLs, compared to 65..70\,MLs for other grain size bins in Model {\f noPD}. Once the first layer has formed, ice mass in Model {\f full} catches up with Model {\f noPD} within 0.2\,Myr because accretion dominates over desorption in the dense core. Table~\ref{tab-mod} shows that the effect of our $E_D$-dependent photodesorption approach is moderate, with H$_2$O:CO:CO$_2$:CH$_3$OH:NH$_3$ final ice abundance ratio being 100:60:39:10:3 in Model {\f noPD} and 100:63:34:10:3 in Model {\f full}.

Fig.~\ref{fig-xcd} shows that changes introduced by chemical desorption are more pronounced than those of photodesorption. The first ice ML in Model {\f noCD} forms already at $t=1.03$\,Myr and on the largest, not smallest grains that is the case in other models. Such a reverse trend can be explained by more efficient hydrogenation of surface O on the lower-temperature large grains along with the fact that H$_2$O synthesis rate is not diminished by an efficient chemical desorption in Model {\f noCD}. The abundance of atomic H on grains is in the vicinity of $10^{-12}$ for all grain size bins in most models. Consequently, the small 0.037\,$\rm \mu$m grains achieve their first ice ML only at 1.08\,Myr and grow the thinnest ice layer of 66\,MLs, while for other grain size bins the end thickness is similar at 72...75\,MLs.

The rapid ice accumulation in Model {\f noCD} means that the first ice ML forms already at $A_V=1.1$\,mag. In Model {\f full} this happens only at $A_V=1.8$\,mag. Observable water ice first appears at 3.2\,mag extinction along the line-of-sight \citep{Whittet01}. Our model is too simple to discern, which of the two mechanisms -- the high-efficiency chemical desorption \citep{Minissale16aa} or the low-efficiency reactive desorption \citep{Garrod06f} -- is more consistent with observations because it depends on a number of factors, such as the history of the cloud , its density, geometry, irradiation intensity, and lifetime of the translucent and dark core stages \citep[see][]{Hocuk16}.

The earlier ice accumulation in Model {\f noCD} starts at $T_d$ in the range of 12...16\,K (a degree higher than in Model {\f full}) and at about 1.5 times higher interstellar irradiation. Both of these aspects promote surface oxidation of CO, resulting in higher initial CO$_2$ ice abundances. At later and colder stages, rapid surface synthesis of H$_2$O wins the competition for surface O and OH because of the inefficiency of the chemical desorption of OH and H$_2$O in hydrogenation reactions in Model {\f noCD}.

Oxygen chemistry in Model {\f noCD} is notably changed by the appearance of surface O$_2$. The early accumulation of O atoms on relatively warm grains in combination with inefficient hydrogenation allows them to combine with the newly formed O$_2$ mostly remaining on the surface \citep[see also][]{Pauly16}. Oxygen ice takes up 3.4\,\% of all oxygen budget, an order of magnitude higher than in Model {\f full}. Abundant O$_2$, together with a 60\,\% higher abundance of H$_2$O$_2$, reduces the abundance of water ice by one fifth. This reduction consequently increases the ratios of carbon oxide ices relative to H$_2$O. The absolute abundances of CO and CO$_2$ have changed only within 7\,\% relative to Model {\f full}.

The H$_2$O:CO:CO$_2$:CH$_3$OH:NH$_3$:O$_2$ final ice abundance ratio in Model {\f noCD} is 100:73:44:10:7:5. This ratio and Fig.~\ref{fig-xcd} reveals another important result -- the lack of an effective chemical desorption allows for the formation of ammonia ice with H$_2$O:NH$_3$ ratios closer to the $\approx$10\,\% value indicated by observations \citep{Boogert11}. Thus, our results predict that the chemical desorption efficiency of nitrogen hydrogenation products NH, NH$_2$, and NH$_3$ should be about an order of magnitude lower than indicated by the method of \citet{Minissale16aa}.

Chemical desorption has also a significant effect in the pseudo-time dependent model {\f const}, where, when the \citet{Minissale16aa} chemical desorption is replaced with $f_{\rm CD}=3$\,\%, the first full ice ML forms already at $t=2$\,kyr, while 90\,\% freeze-out is reached about 0.1\,Myr earlier.

As far as we know, this is the first published study of a multi-grain astrochemical model considering chemical desorption based on \citet{Minissale16aa}. While there are a few other such `firsts', combining multiple grain size bins with chemical desorption is important. In multi-grain models, grain surface area is higher by about a factor of two, allowing for an earlier accretion of ices. This means accretion at higher gas temperatures of around 20\,K, with higher thermal velocities that make accretion even more rapid. Small grains increase their surface area with each adsorbed ice ML, leading to a possibility of a runaway freeze-out. When the chemical desorption is added, it delays the formation of the first ice ML on the bare grain, where it is most effective. Further ice growth continues to be hampered because tens of per cent of hydrogenation reaction products going to the gas phase. Therefore, for multi-grain models considering bulk-ice (meaning that a major part of ice is isolated from desorption) and efficient chemical desorption, a completely different gas-grain dynamics occurs, producing results that can be superficially similar to those of much simpler models.

\section{Summary and conclusions}
\label{cncl}

The inclusion of the variable $E_D$ has mostly insignificant chemical effects in the rapid ice accretion regime of the cold core and a moderate effect in the contracting core model {\ full}. A few species, such as H$_2$O$_2$, CH$_3$OH have their gas-phase abundances elevated by up to an order of magnitude, while the abundances of the whole carbon chain compound class have become lower by about a factor of 2. The first ice ML appears 56\,kyr earlier than in the model {\f noEd} without variable $E_D$. A higher abundance of CO$_2$ is the main effect on ice composition. Such changes are comparable with those introduced by the multi-grain or multi-layer approaches \citep{Acharyya11,Vasyunin13}. Thus, variable $E_D$ is among necessary steps for accurate astrochemical modelling. In this study it has been implemented in a rather careful manner with $E_{D,\rm np}/E_{D,\rm pol}=0.8$, compared to \citet{Bergin97}, who took $E_{D,\rm np}/E_{D,\rm pol}=0.56$.

The inclusion of the chemical desorption \citep{Minissale16aa} in the multi-grain multi-layer gas-surface chemical model turned out to be of major importance. In effect, chemical desorption decreases the rate of icy molecule synthesis, delaying the formation of the first ice ML by almost 0.2\,Myr, as demonstrated by the comparison of Models {\f full} and {\f noCD} in Sect.~\ref{cdpd}. This aspect is not readily apparent in pseudo-time dependent models, such as \citet{Vasyunin17} and \citet{Rawlings21}, and has been largely missed so far.

Variable $E_D$ has its most visible effect on the abundances of major icy species by increasing the amount of CO$_2$ at the expense of CO and H$_2$O. The final calculated relative ice abundances H$_2$O:CO:CO$_2$ were 100:63:34, in agreement with heavily shielded dense cores \citep{Whittet11,Boogert13}. The removal of $E_D$-dependence for one or two molecular-level processes does not change the approximate calculated ice abundance ratio for water and carbon oxides. It indicates a resilience in the modelling results in the sense that they do not depend on a few sensible parameters. The H$_2$O:CO:CO$_2$ ice abundance ratio is determined by the partial combination of co-adsorbing atomic O and CO into CO$_2$ ices \citep{Ruffle01}.

A mechanism that regulates the balance between water and carbon oxide ices has been sought by several studies \citep{Nejad92,Bergin95,Ruffle01,Roberts07,K15apj1}. In this study, such a mechanism is chemical desorption, photodesorption, and desorption by H+H surface reactions, all together. Their general effect is ensuring co-adsorption of some atomic O and CO onto grains sufficiently warm for their partial conversion into CO$_2$ ices \citep{Garrod11}. If one of the desorption mechanisms is ineffective in a given cloud core, others can partially offset it, retaining the characteristic $\rm H_2O>[CO\approx CO_2$] ice abundance sequence (cf. Figs. \ref{fig-epoch} and \ref{fig-xcd}). The latter aspect explains the ubiquity of the H$_2$O:CO:CO$_2$ ice ratio observations \citep{Gibb04,Whittet07}. Our model is unable to explain a separate problem -- the relatively low depletion of CO in interstellar clouds \citep{Leger83,Leger85,Whittet10}. Other mechanisms that can contribute to the balance between water and carbon oxides are photodesorption by infrared photons \citep{Williams92,Dzegilenko95,Santos23} and cosmic-ray induced desorption (Sect.~\ref{mchm}) that does not have a great effect in our model. Grain size distribution plays a role, with smaller grains being warmer and more efficient at producing CO$_2$ ice.

The abundance of solid CH$_3$OH, in combination with desorption facilitated by lower $E_D$, is sufficient to explain its observed gas-phase abundance in dark cores \citep{Bacmann12,Cernicharo12}. This is not true for most other COMs, probably due to a limited reaction network. Regarding NH$_3$, underproduction of ammonia in Model {\f full} might be explained by an over-effective chemical desorption of hydrogen nitrides \citep[corroborated by][]{Sipila19}. The chemistry of COMs and nitrogen both merit detailed investigation with this model.

Summarising, the model combines ten gradual steps in theoretical (surface) astrochemistry: (1) several grain size bins, (2) bulk ice that is (3) chemically active and (4) consists of several separate layers, experiment-based estimates for (5) chemical and (6) photodesorption, updated estimates for (7) cosmic-ray induced desorption and (8) desorption by surface H atom combination reaction heat, (9) a general approach for adjusting $E_D$ on bare grains and (10) a method for estimating $E_D$ in weakly polar icy environment. While any one of these features may not contribute much and its method can be improved, all together they bring new understanding on how interstellar grain surface chemistry operates, as indicated by the previous studies that have investigated many of these features separately (Sect.~\ref{intrd}). Features (6), (9), and (10) are described in a novel way in this study. The above-discussed delay in ice formation occurs because of combining features (5) and (9). The production of CO$_2$ ice is positively affected by (1), (3), and (10), and negatively by (5), (6), (8), and (9).

\begin{acknowledgements}
This research is funded by the Latvian Science Council grant `Desorption of icy molecules in the interstellar medium (DIMD)', project No. lzp-2021/1-0076. JK thanks Ventspils City Council for support. This research has made use of NASA’s Astrophysics Data System. We thank the anonymous referee for necessary improvements.
\end{acknowledgements}

   \bibliographystyle{aa}
   \bibliography{Edes}

\begin{thebibliography}{157}
\expandafter\ifx\csname natexlab\endcsname\relax\def\natexlab#1{#1}\fi

\bibitem[{{Acharyya} {et~al.}(2007){Acharyya}, {Fuchs}, {Fraser}, {van
  Dishoeck}, \& {Linnartz}}]{Acharyya07}
{Acharyya}, K., {Fuchs}, G.~W., {Fraser}, H.~J., {van Dishoeck}, E.~F., \&
  {Linnartz}, H. 2007, \aap, 466, 1005

\bibitem[{{Acharyya} {et~al.}(2011){Acharyya}, {Hassel}, \&
  {Herbst}}]{Acharyya11}
{Acharyya}, K., {Hassel}, G.~E., \& {Herbst}, E. 2011, \apj, 732, 73

\bibitem[{{Ahirwar} {et~al.}(2022){Ahirwar}, {Gurav}, {Gadre}, \&
  {Deshmukh}}]{Ahirwar22}
{Ahirwar}, M.~B., {Gurav}, N.~D., {Gadre}, S.~R., \& {Deshmukh}, M.~M. 2022,
  Physical Chemistry Chemical Physics, 24, 15462

\bibitem[{{Ahirwar} {et~al.}(2021){Ahirwar}, {Patkar}, {Yadav}, \&
  {Deshmukh}}]{Ahirwar21}
{Ahirwar}, M.~B., {Patkar}, D., {Yadav}, I., \& {Deshmukh}, M.~M. 2021,
  Physical Chemistry Chemical Physics, 23, 17224

\bibitem[{{Alkorta} \& {Legon}(2023)}]{Alkorta23}
{Alkorta}, I. \& {Legon}, A. 2023, JPCA, 127, 4715

\bibitem[{{Andersen} {et~al.}(2015){Andersen}, {Heimdal}, \& {Wugt
  Larsen}}]{Andersen15}
{Andersen}, J., {Heimdal}, J., \& {Wugt Larsen}, R. 2015, Physical Chemistry
  Chemical Physics, 17, 23761

\bibitem[{{Andersson} {et~al.}(2006){Andersson}, {Al-Halabi}, {Kroes}, \& {van
  Dishoeck}}]{Andersson06}
{Andersson}, S., {Al-Halabi}, A., {Kroes}, G.-J., \& {van Dishoeck}, E.~F.
  2006, \jcp, 124, 064715

\bibitem[{{Andersson} \& {van Dishoeck}(2008)}]{Andersson08}
{Andersson}, S. \& {van Dishoeck}, E.~F. 2008, \aap, 491, 907

\bibitem[{{Arasa} {et~al.}(2015){Arasa}, {Koning}, {Kroes}, {Walsh}, \& {van
  Dishoeck}}]{Arasa15}
{Arasa}, C., {Koning}, J., {Kroes}, G.-J., {Walsh}, C., \& {van Dishoeck},
  E.~F. 2015, \aap, 575, A121

\bibitem[{{Bacmann} {et~al.}(2012){Bacmann}, {Taquet}, {Faure}, {Kahane}, \&
  {Ceccarelli}}]{Bacmann12}
{Bacmann}, A., {Taquet}, V., {Faure}, A., {Kahane}, C., \& {Ceccarelli}, C.
  2012, \aap, 541, L12

\bibitem[{{Basalg{\`e}te} {et~al.}(2021){Basalg{\`e}te}, {Oca{\~n}a},
  {F{\'e}raud}, {Romanzin}, {Philippe}, {Michaut}, {Fillion}, \&
  {Bertin}}]{Basalgete21}
{Basalg{\`e}te}, R., {Oca{\~n}a}, A.~J., {F{\'e}raud}, G., {et~al.} 2021, \apj,
  922, 213

\bibitem[{{Bergin} \& {Langer}(1997)}]{Bergin97}
{Bergin}, E.~A. \& {Langer}, W.~D. 1997, \apj, 486, 316

\bibitem[{{Bergin} {et~al.}(1995){Bergin}, {Langer}, \& {Goldsmith}}]{Bergin95}
{Bergin}, E.~A., {Langer}, W.~D., \& {Goldsmith}, P.~F. 1995, \apj, 441, 222

\bibitem[{{Bertin} {et~al.}(2012){Bertin}, {Fayolle}, {Romanzin}, {{\"O}berg},
  {Michaut}, {Moudens}, {Philippe}, {Jeseck}, {Linnartz}, \&
  {Fillion}}]{Bertin12}
{Bertin}, M., {Fayolle}, E.~C., {Romanzin}, C., {et~al.} 2012, Physical
  Chemistry Chemical Physics, 14, 9929

\bibitem[{{Bertin} {et~al.}(2013){Bertin}, {Fayolle}, {Romanzin}, {Poderoso},
  {Michaut}, {Philippe}, {Jeseck}, {{\"O}berg}, {Linnartz}, \&
  {Fillion}}]{Bertin13}
{Bertin}, M., {Fayolle}, E.~C., {Romanzin}, C., {et~al.} 2013, \apj, 779, 120

\bibitem[{{Bertin} {et~al.}(2016){Bertin}, {Romanzin}, {Doronin}, {Philippe},
  {Jeseck}, {Ligterink}, {Linnartz}, {Michaut}, \& {Fillion}}]{Bertin16}
{Bertin}, M., {Romanzin}, C., {Doronin}, M., {et~al.} 2016, \apjl, 817, L12

\bibitem[{{Boogert} {et~al.}(2013){Boogert}, {Chiar}, {Knez}, {{\"O}berg},
  {Mundy}, {Pendleton}, {Tielens}, \& {van Dishoeck}}]{Boogert13}
{Boogert}, A.~C.~A., {Chiar}, J.~E., {Knez}, C., {et~al.} 2013, \apj, 777, 73

\bibitem[{{Boogert} {et~al.}(2011){Boogert}, {Huard}, {Cook}, {Chiar}, {Knez},
  {Decin}, {Blake}, {Tielens}, \& {van Dishoeck}}]{Boogert11}
{Boogert}, A.~C.~A., {Huard}, T.~L., {Cook}, A.~M., {et~al.} 2011, \apj, 729,
  92

\bibitem[{{Boryskina} {et~al.}(2007){Boryskina}, {Bolbukh}, {Semenov}, {Gasan},
  \& {Maleev}}]{Boryskina07}
{Boryskina}, O.~P., {Bolbukh}, T.~V., {Semenov}, M.~A., {Gasan}, A.~I., \&
  {Maleev}, V.~Y. 2007, Journal of Molecular Structure, 827, 1

\bibitem[{{Bredeh{\"o}ft}(2020)}]{Bredehoft20}
{Bredeh{\"o}ft}, J.~H. 2020, Frontiers in Astronomy and Space Sciences, 7, 33

\bibitem[{{Brown} {et~al.}(1988){Brown}, {Charnley}, \& {Millar}}]{Brown88}
{Brown}, P.~D., {Charnley}, S.~B., \& {Millar}, T.~J. 1988, \mnras, 231, 409

\bibitem[{{Bulak} {et~al.}(2020){Bulak}, {Paardekooper}, {Fedoseev}, \&
  {Linnartz}}]{Bulak20}
{Bulak}, M., {Paardekooper}, D.~M., {Fedoseev}, G., \& {Linnartz}, H. 2020,
  \aap, 636, A32

\bibitem[{{Bulak} {et~al.}(2023){Bulak}, {Paardekooper}, {Fedoseev}, {Samarth},
  \& {Linnartz}}]{Bulak23}
{Bulak}, M., {Paardekooper}, D.~M., {Fedoseev}, G., {Samarth}, P., \&
  {Linnartz}, H. 2023, \aap, 677, A99

\bibitem[{{Carrascosa} {et~al.}(2019){Carrascosa}, {Hsiao}, {Sie}, {Mu{\~n}oz
  Caro}, \& {Chen}}]{Carrascosa19}
{Carrascosa}, H., {Hsiao}, L.~C., {Sie}, N.~E., {Mu{\~n}oz Caro}, G.~M., \&
  {Chen}, Y.~J. 2019, \mnras, 486, 1985

\bibitem[{{Cazaux} {et~al.}(2010){Cazaux}, {Cobut}, {Marseille}, {Spaans}, \&
  {Caselli}}]{Cazaux10}
{Cazaux}, S., {Cobut}, V., {Marseille}, M., {Spaans}, M., \& {Caselli}, P.
  2010, \aap, 522, A74

\bibitem[{{Cernicharo} {et~al.}(2012){Cernicharo}, {Marcelino}, {Roueff},
  {Gerin}, {Jim{\'e}nez-Escobar}, \& {Mu{\~n}oz Caro}}]{Cernicharo12}
{Cernicharo}, J., {Marcelino}, N., {Roueff}, E., {et~al.} 2012, \apjl, 759, L43

\bibitem[{{Chaabouni} {et~al.}(2018){Chaabouni}, {Diana}, {Nguyen}, \&
  {Dulieu}}]{Chaabouni18}
{Chaabouni}, H., {Diana}, S., {Nguyen}, T., \& {Dulieu}, F. 2018, \aap, 612,
  A47

\bibitem[{{Chaabouni} {et~al.}(2012){Chaabouni}, {Minissale}, {Manic{\`o}},
  {Congiu}, {Noble}, {Baouche}, {Accolla}, {Lemaire}, {Pirronello}, \&
  {Dulieu}}]{Chaabouni12}
{Chaabouni}, H., {Minissale}, M., {Manic{\`o}}, G., {et~al.} 2012, \jcp, 137,
  234706

\bibitem[{{Chang} {et~al.}(2007){Chang}, {Cuppen}, \& {Herbst}}]{Chang07}
{Chang}, Q., {Cuppen}, H.~M., \& {Herbst}, E. 2007, \aap, 469, 973

\bibitem[{{Chang} \& {Herbst}(2014)}]{Chang14}
{Chang}, Q. \& {Herbst}, E. 2014, \apj, 787, 135

\bibitem[{{Chang} \& {Herbst}(2016)}]{Chang16}
{Chang}, Q. \& {Herbst}, E. 2016, \apj, 819, 145

\bibitem[{{Chen} {et~al.}(2018){Chen}, {Chang}, \& {Xi}}]{Chen18}
{Chen}, L.-F., {Chang}, Q., \& {Xi}, H.-W. 2018, \mnras, 479, 2988

\bibitem[{{Chen} {et~al.}(2014){Chen}, {Chuang}, {Mu{\~n}oz Caro}, {Nuevo},
  {Chu}, {Yih}, {Ip}, \& {Wu}}]{Chen14}
{Chen}, Y.~J., {Chuang}, K.~J., {Mu{\~n}oz Caro}, G.~M., {et~al.} 2014, \apj,
  781, 15

\bibitem[{{Collings} {et~al.}(2004){Collings}, {Anderson}, {Chen}, {Dever},
  {Viti}, {Williams}, \& {McCoustra}}]{Collings04}
{Collings}, M.~P., {Anderson}, M.~A., {Chen}, R., {et~al.} 2004, \mnras, 354,
  1133

\bibitem[{{Cruz-Diaz} {et~al.}(2018){Cruz-Diaz}, {Mart{\'\i}n-Dom{\'e}nech},
  {Moreno}, {Mu{\~n}oz Caro}, \& {Chen}}]{Cruz18}
{Cruz-Diaz}, G.~A., {Mart{\'\i}n-Dom{\'e}nech}, R., {Moreno}, E., {Mu{\~n}oz
  Caro}, G.~M., \& {Chen}, Y.-J. 2018, \mnras, 474, 3080

\bibitem[{{Cruz-Diaz} {et~al.}(2016){Cruz-Diaz}, {Mart{\'\i}n-Dom{\'e}nech},
  {Mu{\~n}oz Caro}, \& {Chen}}]{Cruz16}
{Cruz-Diaz}, G.~A., {Mart{\'\i}n-Dom{\'e}nech}, R., {Mu{\~n}oz Caro}, G.~M., \&
  {Chen}, Y.~J. 2016, \aap, 592, A68

\bibitem[{{Cuppen} \& {Herbst}(2007)}]{Cuppen07}
{Cuppen}, H.~M. \& {Herbst}, E. 2007, \apj, 668, 294

\bibitem[{{Das} {et~al.}(2018){Das}, {Sil}, {Gorai}, {Chakrabarti}, \&
  {Loison}}]{Das18}
{Das}, A., {Sil}, M., {Gorai}, P., {Chakrabarti}, S.~K., \& {Loison}, J.~C.
  2018, \apjs, 237, 9

\bibitem[{{Doronin} {et~al.}(2015){Doronin}, {Bertin}, {Michaut}, {Philippe},
  \& {Fillion}}]{Doronin15}
{Doronin}, M., {Bertin}, M., {Michaut}, X., {Philippe}, L., \& {Fillion}, J.~H.
  2015, \jcp, 143, 084703

\bibitem[{{Duley} \& {Williams}(1993)}]{Duley93}
{Duley}, W.~W. \& {Williams}, D.~A. 1993, \mnras, 260, 37

\bibitem[{{Dulieu} {et~al.}(2013){Dulieu}, {Congiu}, {Noble}, {Baouche},
  {Chaabouni}, {Moudens}, {Minissale}, \& {Cazaux}}]{Dulieu13}
{Dulieu}, F., {Congiu}, E., {Noble}, J., {et~al.} 2013, Scientific Reports, 3,
  1338

\bibitem[{{Dupuy} {et~al.}(2017{\natexlab{a}}){Dupuy}, {Bertin}, {F{\'e}raud},
  {Michaut}, {Jeseck}, {Doronin}, {Philippe}, {Romanzin}, \&
  {Fillion}}]{Dupuy17c}
{Dupuy}, R., {Bertin}, M., {F{\'e}raud}, G., {et~al.} 2017{\natexlab{a}}, \aap,
  603, A61

\bibitem[{{Dupuy} {et~al.}(2017{\natexlab{b}}){Dupuy}, {F{\'e}raud}, {Bertin},
  {Michaut}, {Putaud}, {Jeseck}, {Philippe}, {Romanzin}, {Baglin}, {Cimino}, \&
  {Fillion}}]{Dupuy17n}
{Dupuy}, R., {F{\'e}raud}, G., {Bertin}, M., {et~al.} 2017{\natexlab{b}}, \aap,
  606, L9

\bibitem[{{Dzegilenko} \& {Herbst}(1995)}]{Dzegilenko95}
{Dzegilenko}, F. \& {Herbst}, E. 1995, \apjl, 443, L81

\bibitem[{{Fayolle} {et~al.}(2016){Fayolle}, {Balfe}, {Loomis}, {Bergner},
  {Graninger}, {Rajappan}, \& {{\"O}berg}}]{Fayolle16}
{Fayolle}, E.~C., {Balfe}, J., {Loomis}, R., {et~al.} 2016, \apjl, 816, L28

\bibitem[{{Fayolle} {et~al.}(2011){Fayolle}, {Bertin}, {Romanzin}, {Michaut},
  {{\"O}berg}, {Linnartz}, \& {Fillion}}]{Fayolle11}
{Fayolle}, E.~C., {Bertin}, M., {Romanzin}, C., {et~al.} 2011, \apjl, 739, L36

\bibitem[{{Fayolle} {et~al.}(2013){Fayolle}, {Bertin}, {Romanzin}, {Poderoso},
  {Philippe}, {Michaut}, {Jeseck}, {Linnartz}, {{\"O}berg}, \&
  {Fillion}}]{Fayolle13}
{Fayolle}, E.~C., {Bertin}, M., {Romanzin}, C., {et~al.} 2013, \aap, 556, A122

\bibitem[{{F{\'e}raud} {et~al.}(2019){F{\'e}raud}, {Bertin}, {Romanzin},
  {Dupuy}, {Le Petit}, {Roueff}, {Philippe}, {Michaut}, {Jeseck}, \&
  {Fillion}}]{Feraud19}
{F{\'e}raud}, G., {Bertin}, M., {Romanzin}, C., {et~al.} 2019, ACS Earth and
  Space Chemistry, 3, 1135

\bibitem[{{Fillion} {et~al.}(2022){Fillion}, {Dupuy}, {F{\'e}raud}, {Romanzin},
  {Philippe}, {Putaud}, {Baglin}, {Cimino}, {Marie-Jeanne}, {Jeseck},
  {Michaut}, \& {Bertin}}]{Fillion22}
{Fillion}, J.-H., {Dupuy}, R., {F{\'e}raud}, G., {et~al.} 2022, ACS Earth and
  Space Chemistry, 6, 100

\bibitem[{{Fillion} {et~al.}(2014){Fillion}, {Fayolle}, {Michaut}, {Doronin},
  {Philippe}, {Rakovski}, {Romanzin}, {Champion}, {{\"O}berg}, {Linnartz}, \&
  {Bertin}}]{Fillion14}
{Fillion}, J.-H., {Fayolle}, E.~C., {Michaut}, X., {et~al.} 2014, Faraday
  Discussions, 168, 533

\bibitem[{{Flynn}(2020)}]{Flynn20}
{Flynn}, G.~J. 2020, in Oxford Research Encyclopedia of Planetary Science
  (Oxford University Press), 143

\bibitem[{{Fredon} {et~al.}(2017){Fredon}, {Lamberts}, \& {Cuppen}}]{Fredon17}
{Fredon}, A., {Lamberts}, T., \& {Cuppen}, H.~M. 2017, \apj, 849, 125

\bibitem[{{Furuya} {et~al.}(2017){Furuya}, {Drozdovskaya}, {Visser}, {van
  Dishoeck}, {Walsh}, {Harsono}, {Hincelin}, \& {Taquet}}]{Furuya17}
{Furuya}, K., {Drozdovskaya}, M.~N., {Visser}, R., {et~al.} 2017, \aap, 599,
  A40

\bibitem[{{Garrod} {et~al.}(2006){Garrod}, {Park}, {Caselli}, \&
  {Herbst}}]{Garrod06f}
{Garrod}, R., {Park}, I.~H., {Caselli}, P., \& {Herbst}, E. 2006, Faraday
  Discussions, 133, 51

\bibitem[{{Garrod}(2013{\natexlab{a}})}]{Garrod13}
{Garrod}, R.~T. 2013{\natexlab{a}}, \apj, 765, 60

\bibitem[{{Garrod}(2013{\natexlab{b}})}]{Garrod13mc}
{Garrod}, R.~T. 2013{\natexlab{b}}, \apj, 778, 158

\bibitem[{{Garrod} \& {Herbst}(2006)}]{Garrod06}
{Garrod}, R.~T. \& {Herbst}, E. 2006, \aap, 457, 927

\bibitem[{{Garrod} {et~al.}(2022){Garrod}, {Jin}, {Matis}, {Jones}, {Willis},
  \& {Herbst}}]{Garrod22}
{Garrod}, R.~T., {Jin}, M., {Matis}, K.~A., {et~al.} 2022, \apjs, 259, 1

\bibitem[{{Garrod} \& {Pauly}(2011)}]{Garrod11}
{Garrod}, R.~T. \& {Pauly}, T. 2011, \apj, 735, 15

\bibitem[{{Garrod} {et~al.}(2008){Garrod}, {Weaver}, \& {Herbst}}]{Garrod08}
{Garrod}, R.~T., {Weaver}, S.~L.~W., \& {Herbst}, E. 2008, ApJ, 682, 283

\bibitem[{{Gavino} {et~al.}(2021){Gavino}, {Dutrey}, {Wakelam}, {Guilloteau},
  {Kobus}, {Wolf}, {Iqbal}, {Di Folco}, {Chapillon}, \& {Pi{\'e}tu}}]{Gavino21}
{Gavino}, S., {Dutrey}, A., {Wakelam}, V., {et~al.} 2021, \aap, 654, A65

\bibitem[{{Ge} {et~al.}(2016){Ge}, {He}, \& {Li}}]{Ge16}
{Ge}, J.~X., {He}, J.~H., \& {Li}, A. 2016, \mnras, 460, L50

\bibitem[{{Gibb} {et~al.}(2004){Gibb}, {Whittet}, {Boogert}, \&
  {Tielens}}]{Gibb04}
{Gibb}, E.~L., {Whittet}, D.~C.~B., {Boogert}, A.~C.~A., \& {Tielens},
  A.~G.~G.~M. 2004, \apjs, 151, 35

\bibitem[{{Gonz{\'a}lez D{\'\i}az} {et~al.}(2019){Gonz{\'a}lez D{\'\i}az},
  {Carrascosa de Lucas}, {Aparicio}, {Mu{\~n}oz Caro}, {Sie}, {Hsiao},
  {Cazaux}, \& {Chen}}]{Gonzalez19}
{Gonz{\'a}lez D{\'\i}az}, C., {Carrascosa de Lucas}, H., {Aparicio}, S.,
  {et~al.} 2019, \mnras, 486, 5519

\bibitem[{{Grassi} {et~al.}(2020){Grassi}, {Bovino}, {Caselli}, {Bovolenta},
  {Vogt-Geisse}, \& {Ercolano}}]{Grassi20}
{Grassi}, T., {Bovino}, S., {Caselli}, P., {et~al.} 2020, \aap, 643, A155

\bibitem[{{Hasegawa} \& {Herbst}(1993{\natexlab{a}})}]{Hasegawa93cr}
{Hasegawa}, T.~I. \& {Herbst}, E. 1993{\natexlab{a}}, \mnras, 261, 83

\bibitem[{{Hasegawa} \& {Herbst}(1993{\natexlab{b}})}]{Hasegawa93m}
{Hasegawa}, T.~I. \& {Herbst}, E. 1993{\natexlab{b}}, \mnras, 263, 589

\bibitem[{{He} {et~al.}(2016){He}, {Acharyya}, \& {Vidali}}]{He16}
{He}, J., {Acharyya}, K., \& {Vidali}, G. 2016, \apj, 825, 89

\bibitem[{{He} {et~al.}(2018){He}, {Emtiaz}, \& {Vidali}}]{He18}
{He}, J., {Emtiaz}, S., \& {Vidali}, G. 2018, \apj, 863, 156

\bibitem[{{Hincelin} {et~al.}(2015){Hincelin}, {Chang}, \&
  {Herbst}}]{Hincelin15}
{Hincelin}, U., {Chang}, Q., \& {Herbst}, E. 2015, \aap, 574, A24

\bibitem[{{Hocuk} \& {Cazaux}(2015)}]{Hocuk15}
{Hocuk}, S. \& {Cazaux}, S. 2015, \aap, 576, A49

\bibitem[{{Hocuk} {et~al.}(2016){Hocuk}, {Cazaux}, {Spaans}, \&
  {Caselli}}]{Hocuk16}
{Hocuk}, S., {Cazaux}, S., {Spaans}, M., \& {Caselli}, P. 2016, \mnras, 456,
  2586

\bibitem[{{Hocuk} {et~al.}(2017){Hocuk}, {Sz{\H{u}}cs}, {Caselli}, {Cazaux},
  {Spaans}, \& {Esplugues}}]{Hocuk17}
{Hocuk}, S., {Sz{\H{u}}cs}, L., {Caselli}, P., {et~al.} 2017, \aap, 604, A58

\bibitem[{{Iqbal} \& {Wakelam}(2018)}]{Iqbal18}
{Iqbal}, W. \& {Wakelam}, V. 2018, \aap, 615, A20

\bibitem[{{Ivlev} {et~al.}(2015){Ivlev}, {Padovani}, {Galli}, \&
  {Caselli}}]{Ivlev15}
{Ivlev}, A.~V., {Padovani}, M., {Galli}, D., \& {Caselli}, P. 2015, \apj, 812,
  135

\bibitem[{{Jin} \& {Garrod}(2020)}]{Jin20}
{Jin}, M. \& {Garrod}, R.~T. 2020, \apjs, 249, 26

\bibitem[{{Johnson} {et~al.}(2000){Johnson}, {Blitz}, \& {Seakins}}]{Johnson00}
{Johnson}, D.~G., {Blitz}, M.~A., \& {Seakins}, P.~W. 2000, Physical Chemistry
  Chemical Physics, 2, 2549

\bibitem[{{Kakkenpara Suresh} {et~al.}(2024){Kakkenpara Suresh}, {Dulieu},
  {Vitorino}, \& {Caselli}}]{Kakkenpara24}
{Kakkenpara Suresh}, S., {Dulieu}, F., {Vitorino}, J., \& {Caselli}, P. 2024,
  \aap, 682, A163

\bibitem[{{Kalv{\= a}ns}(2015)}]{K15apj1}
{Kalv{\= a}ns}, J. 2015, \apj, 803, 52

\bibitem[{{Kalv{\= a}ns}(2018)}]{K18mn}
{Kalv{\= a}ns}, J. 2018, \mnras, 478, 2753

\bibitem[{{Kalv{\= a}ns} \& {Kalnin}(2019)}]{KK19}
{Kalv{\= a}ns}, J. \& {Kalnin}, J.~R. 2019, \mnras, 486, 2050

\bibitem[{{Kalv{\= a}ns} \& {Shmeld}(2010)}]{KS10}
{Kalv{\= a}ns}, J. \& {Shmeld}, I. 2010, \aap, 521, A37

\bibitem[{{Kalv{\={a}}ns}(2015)}]{K15apj2}
{Kalv{\={a}}ns}, J. 2015, \apj, 806, 196

\bibitem[{{Kalv{\={a}}ns}(2021)}]{K21}
{Kalv{\={a}}ns}, J. 2021, \apj, 910, 54

\bibitem[{{Kalv{\={a}}ns} \& {Kalnin}(2022)}]{KK22}
{Kalv{\={a}}ns}, J. \& {Kalnin}, J.~R. 2022, \apjs, 263, 5

\bibitem[{{Kalv{\={a}}ns} \& {Silsbee}(2022)}]{KS22}
{Kalv{\={a}}ns}, J. \& {Silsbee}, K. 2022, \mnras, 515, 785

\bibitem[{{Kikuta} {et~al.}(2008){Kikuta}, {Ishimoto}, \&
  {Nagashima}}]{Kikuta08}
{Kikuta}, Y., {Ishimoto}, T., \& {Nagashima}, U. 2008, Chemical Physics, 354,
  218

\bibitem[{{Leger}(1983)}]{Leger83}
{Leger}, A. 1983, \aap, 123, 271

\bibitem[{{Leger} {et~al.}(1985){Leger}, {Jura}, \& {Omont}}]{Leger85}
{Leger}, A., {Jura}, M., \& {Omont}, A. 1985, \aap, 144, 147

\bibitem[{{Mart{\'\i}n-Dom{\'e}nech} {et~al.}(2018){Mart{\'\i}n-Dom{\'e}nech},
  {Cruz-D{\'\i}az}, \& {Mu{\~n}oz Caro}}]{Martin18}
{Mart{\'\i}n-Dom{\'e}nech}, R., {Cruz-D{\'\i}az}, G.~A., \& {Mu{\~n}oz Caro},
  G.~M. 2018, \mnras, 473, 2575

\bibitem[{{Mart{\'\i}n-Dom{\'e}nech} {et~al.}(2015){Mart{\'\i}n-Dom{\'e}nech},
  {Manzano-Santamar{\'\i}a}, {Mu{\~n}oz Caro}, {Cruz-D{\'\i}az}, {Chen},
  {Herrero}, \& {Tanarro}}]{Martin15}
{Mart{\'\i}n-Dom{\'e}nech}, R., {Manzano-Santamar{\'\i}a}, J., {Mu{\~n}oz
  Caro}, G.~M., {et~al.} 2015, \aap, 584, A14

\bibitem[{{Mart{\'\i}n-Dom{\'e}nech} {et~al.}(2014){Mart{\'\i}n-Dom{\'e}nech},
  {Mu{\~n}oz Caro}, {Bueno}, \& {Goesmann}}]{Martin14}
{Mart{\'\i}n-Dom{\'e}nech}, R., {Mu{\~n}oz Caro}, G.~M., {Bueno}, J., \&
  {Goesmann}, F. 2014, \aap, 564, A8

\bibitem[{{Mathis} {et~al.}(1977){Mathis}, {Rumpl}, \& {Nordsieck}}]{Mathis77}
{Mathis}, J.~S., {Rumpl}, W., \& {Nordsieck}, K.~H. 1977, \apj, 217, 425

\bibitem[{{McClure} {et~al.}(2023){McClure}, {Rocha}, {Pontoppidan}, {Crouzet},
  {Chu}, {Dartois}, {Lamberts}, {Noble}, {Pendleton}, {Perotti}, {Qasim},
  {Rachid}, {Smith}, {Sun}, {Beck}, {Boogert}, {Brown}, {Caselli}, {Charnley},
  {Cuppen}, {Dickinson}, {Drozdovskaya}, {Egami}, {Erkal}, {Fraser}, {Garrod},
  {Harsono}, {Ioppolo}, {Jim{\'e}nez-Serra}, {Jin}, {J{\o}rgensen},
  {Kristensen}, {Lis}, {McCoustra}, {McGuire}, {Melnick}, {{\~A}-berg},
  {Palumbo}, {Shimonishi}, {Sturm}, {van Dishoeck}, \& {Linnartz}}]{McClure23}
{McClure}, M.~K., {Rocha}, W.~R.~M., {Pontoppidan}, K.~M., {et~al.} 2023,
  Nature Astronomy, 7, 431

\bibitem[{{McElroy} {et~al.}(2013){McElroy}, {Walsh}, {Markwick}, {Cordiner},
  {Smith}, \& {Millar}}]{McElroy13}
{McElroy}, D., {Walsh}, C., {Markwick}, A.~J., {et~al.} 2013, \aap, 550, A36

\bibitem[{{Minissale} \& {Dulieu}(2014)}]{Minissale14}
{Minissale}, M. \& {Dulieu}, F. 2014, \jcp, 141, 014304

\bibitem[{{Minissale} {et~al.}(2016{\natexlab{a}}){Minissale}, {Dulieu},
  {Cazaux}, \& {Hocuk}}]{Minissale16aa}
{Minissale}, M., {Dulieu}, F., {Cazaux}, S., \& {Hocuk}, S. 2016{\natexlab{a}},
  \aap, 585, A24

\bibitem[{{Minissale} {et~al.}(2016{\natexlab{b}}){Minissale}, {Moudens},
  {Baouche}, {Chaabouni}, \& {Dulieu}}]{Minissale16mn}
{Minissale}, M., {Moudens}, A., {Baouche}, S., {Chaabouni}, H., \& {Dulieu}, F.
  2016{\natexlab{b}}, \mnras, 458, 2953

\bibitem[{{Molpeceres} {et~al.}(2023){Molpeceres}, {Zaverkin}, {Furuya},
  {Aikawa}, \& {K{\"a}stner}}]{Molpeceres23}
{Molpeceres}, G., {Zaverkin}, V., {Furuya}, K., {Aikawa}, Y., \& {K{\"a}stner},
  J. 2023, \aap, 673, A51

\bibitem[{{Mu{\~n}oz Caro} {et~al.}(2016){Mu{\~n}oz Caro}, {Chen}, {Aparicio},
  {Jim{\'e}nez-Escobar}, {Rosu-Finsen}, {Lasne}, \& {McCoustra}}]{Munoz16}
{Mu{\~n}oz Caro}, G.~M., {Chen}, Y.~J., {Aparicio}, S., {et~al.} 2016, \aap,
  589, A19

\bibitem[{{Mu{\~n}oz Caro} {et~al.}(2010){Mu{\~n}oz Caro},
  {Jim{\'e}nez-Escobar}, {Mart{\'\i}n-Gago}, {Rogero}, {Atienza}, {Puertas},
  {Sobrado}, \& {Torres-Redondo}}]{Munoz10}
{Mu{\~n}oz Caro}, G.~M., {Jim{\'e}nez-Escobar}, A., {Mart{\'\i}n-Gago},
  J.~{\'A}., {et~al.} 2010, \aap, 522, A108

\bibitem[{{Nejad} \& {Williams}(1992)}]{Nejad92}
{Nejad}, L.~A.~M. \& {Williams}, D.~A. 1992, \mnras, 255, 441

\bibitem[{{Noble} {et~al.}(2012){Noble}, {Congiu}, {Dulieu}, \&
  {Fraser}}]{Noble12}
{Noble}, J.~A., {Congiu}, E., {Dulieu}, F., \& {Fraser}, H.~J. 2012, \mnras,
  421, 768

\bibitem[{{Oba} {et~al.}(2018){Oba}, {Tomaru}, {Lamberts}, {Kouchi}, \&
  {Watanabe}}]{Oba18}
{Oba}, Y., {Tomaru}, T., {Lamberts}, T., {Kouchi}, A., \& {Watanabe}, N. 2018,
  Nature Astronomy, 2, 228

\bibitem[{{{\"O}berg} {et~al.}(2009{\natexlab{a}}){{\"O}berg}, {Linnartz},
  {Visser}, \& {van Dishoeck}}]{Oberg09h2o}
{{\"O}berg}, K.~I., {Linnartz}, H., {Visser}, R., \& {van Dishoeck}, E.~F.
  2009{\natexlab{a}}, \apj, 693, 1209

\bibitem[{{{\"O}berg} {et~al.}(2005){{\"O}berg}, {van Broekhuizen}, {Fraser},
  {Bisschop}, {van Dishoeck}, \& {Schlemmer}}]{Oberg05}
{{\"O}berg}, K.~I., {van Broekhuizen}, F., {Fraser}, H.~J., {et~al.} 2005,
  \apjl, 621, L33

\bibitem[{{{\"O}berg} {et~al.}(2009{\natexlab{b}}){{\"O}berg}, {van Dishoeck},
  \& {Linnartz}}]{Oberg09co}
{{\"O}berg}, K.~I., {van Dishoeck}, E.~F., \& {Linnartz}, H.
  2009{\natexlab{b}}, \aap, 496, 281

\bibitem[{{Paardekooper} {et~al.}(2016){Paardekooper}, {Fedoseev}, {Riedo}, \&
  {Linnartz}}]{Paardekooper16co}
{Paardekooper}, D.~M., {Fedoseev}, G., {Riedo}, A., \& {Linnartz}, H. 2016,
  \aap, 596, A72

\bibitem[{{Padovani} {et~al.}(2009){Padovani}, {Galli}, \&
  {Glassgold}}]{Padovani09}
{Padovani}, M., {Galli}, D., \& {Glassgold}, A.~E. 2009, \aap, 501, 619

\bibitem[{{Pantaleone} {et~al.}(2021){Pantaleone}, {Enrique-Romero},
  {Ceccarelli}, {Ferrero}, {Balucani}, {Rimola}, \& {Ugliengo}}]{Pantaleone21}
{Pantaleone}, S., {Enrique-Romero}, J., {Ceccarelli}, C., {et~al.} 2021, \apj,
  917, 49

\bibitem[{{Pantaleone} {et~al.}(2020){Pantaleone}, {Enrique-Romero},
  {Ceccarelli}, {Ugliengo}, {Balucani}, \& {Rimola}}]{Pantaleone20}
{Pantaleone}, S., {Enrique-Romero}, J., {Ceccarelli}, C., {et~al.} 2020, \apj,
  897, 56

\bibitem[{{Pauly} \& {Garrod}(2016)}]{Pauly16}
{Pauly}, T. \& {Garrod}, R.~T. 2016, \apj, 817, 146

\bibitem[{{Pauly} \& {Garrod}(2018)}]{Pauly18}
{Pauly}, T. \& {Garrod}, R.~T. 2018, \apj, 854, 13

\bibitem[{{Pavlyuchenkov} \& {Zhilkin}(2013)}]{Pavly13}
{Pavlyuchenkov}, Y.~N. \& {Zhilkin}, A.~G. 2013, Astronomy Reports, 57, 641

\bibitem[{{Pavlyuchenkov} {et~al.}(2015){Pavlyuchenkov}, {Zhilkin}, {Vorobyov},
  \& {Fateeva}}]{Pavly15}
{Pavlyuchenkov}, Y.~N., {Zhilkin}, A.~G., {Vorobyov}, E.~I., \& {Fateeva},
  A.~M. 2015, Astronomy Reports, 59, 133

\bibitem[{{Penteado} {et~al.}(2017){Penteado}, {Walsh}, \&
  {Cuppen}}]{Penteado17}
{Penteado}, E.~M., {Walsh}, C., \& {Cuppen}, H.~M. 2017, \apj, 844, 71

\bibitem[{{Potapov} {et~al.}(2019){Potapov}, {J{\"a}ger}, \&
  {Henning}}]{Potapov19}
{Potapov}, A., {J{\"a}ger}, C., \& {Henning}, T. 2019, \apj, 880, 12

\bibitem[{{Rawlings}(2022)}]{Rawlings22}
{Rawlings}, J. M.~C. 2022, \mnras, 517, 3804

\bibitem[{{Rawlings} \& {Williams}(2021)}]{Rawlings21}
{Rawlings}, J. M.~C. \& {Williams}, D.~A. 2021, \mnras, 500, 5117

\bibitem[{{Roberts} {et~al.}(2007){Roberts}, {Rawlings}, {Viti}, \&
  {Williams}}]{Roberts07}
{Roberts}, J.~F., {Rawlings}, J.~M.~C., {Viti}, S., \& {Williams}, D.~A. 2007,
  \mnras, 382, 733

\bibitem[{{Ruffle} \& {Herbst}(2001)}]{Ruffle01}
{Ruffle}, D.~P. \& {Herbst}, E. 2001, \mnras, 324, 1054

\bibitem[{{Sandford} \& {Allamandola}(1988)}]{Sandford88}
{Sandford}, S.~A. \& {Allamandola}, L.~J. 1988, \icarus, 76, 201

\bibitem[{{Santos} {et~al.}(2023){Santos}, {Chuang}, {Schrauwen}, {Traspas
  Mui{\~n}a}, {Zhang}, {Cuppen}, {Redlich}, {Linnartz}, \&
  {Ioppolo}}]{Santos23}
{Santos}, J.~C., {Chuang}, K.~J., {Schrauwen}, J.~G.~M., {et~al.} 2023, \aap,
  672, A112

\bibitem[{{Semenov} {et~al.}(2010){Semenov}, {Hersant}, {Wakelam}, {Dutrey},
  {Chapillon}, {Guilloteau}, {Henning}, {Launhardt}, {Pi{\'e}tu}, \&
  {Schreyer}}]{Semenov10}
{Semenov}, D., {Hersant}, F., {Wakelam}, V., {et~al.} 2010, \aap, 522, A42

\bibitem[{{Shingledecker} {et~al.}(2019){Shingledecker}, {Vasyunin}, {Herbst},
  \& {Caselli}}]{Shingle19}
{Shingledecker}, C.~N., {Vasyunin}, A., {Herbst}, E., \& {Caselli}, P. 2019,
  \apj, 876, 140

\bibitem[{Shinoda(1969)}]{Shinoda69}
Shinoda, T. 1969, Bulletin of the Chemical Society of Japan, 42, 2815

\bibitem[{{Sie} {et~al.}(2022){Sie}, {Cho}, {Huang}, {Mu{\~n}oz Caro}, {Hsiao},
  {Lin}, \& {Chen}}]{Sie22}
{Sie}, N.-E., {Cho}, Y.-T., {Huang}, C.-H., {et~al.} 2022, \apj, 938, 48

\bibitem[{{Silsbee} {et~al.}(2021){Silsbee}, {Caselli}, \& {Ivlev}}]{Silsbee21}
{Silsbee}, K., {Caselli}, P., \& {Ivlev}, A.~V. 2021, \mnras, 507, 6205

\bibitem[{{Silsbee} {et~al.}(2020){Silsbee}, {Ivlev}, {Sipil{\"a}}, {Caselli},
  \& {Zhao}}]{Silsbee20}
{Silsbee}, K., {Ivlev}, A.~V., {Sipil{\"a}}, O., {Caselli}, P., \& {Zhao}, B.
  2020, \aap, 641, A39

\bibitem[{{Sipil{\"a}} {et~al.}(2019){Sipil{\"a}}, {Caselli}, {Redaelli},
  {Juvela}, \& {Bizzocchi}}]{Sipila19}
{Sipil{\"a}}, O., {Caselli}, P., {Redaelli}, E., {Juvela}, M., \& {Bizzocchi},
  L. 2019, \mnras, 487, 1269

\bibitem[{{Sipil{\"a}} {et~al.}(2020){Sipil{\"a}}, {Zhao}, \&
  {Caselli}}]{Sipila20}
{Sipil{\"a}}, O., {Zhao}, B., \& {Caselli}, P. 2020, \aap, 640, A94

\bibitem[{{Spanu} {et~al.}(2008){Spanu}, {Sterpone}, {Ferraro}, {Sorella}, \&
  {Guidoni}}]{Spanu08}
{Spanu}, L., {Sterpone}, F., {Ferraro}, L., {Sorella}, S., \& {Guidoni}, L.
  2008, in APS March Meeting Abstracts, APS Meeting Abstracts, Q26.006

\bibitem[{{Sterpone} {et~al.}(2008){Sterpone}, {Spanu}, {Ferraro}, {Sorella},
  \& {Guidoni}}]{Sterpone08}
{Sterpone}, F., {Spanu}, L., {Ferraro}, L., {Sorella}, S., \& {Guidoni}, L.
  2008, arXiv e-prints, arXiv:0806.4169

\bibitem[{{Takahashi} \& {Williams}(2000)}]{Takahashi00}
{Takahashi}, J. \& {Williams}, D.~A. 2000, \mnras, 314, 273

\bibitem[{{Taquet} {et~al.}(2012){Taquet}, {Ceccarelli}, \&
  {Kahane}}]{Taquet12}
{Taquet}, V., {Ceccarelli}, C., \& {Kahane}, C. 2012, \aap, 538, A42

\bibitem[{{Taquet} {et~al.}(2014){Taquet}, {Charnley}, \&
  {Sipil{\"a}}}]{Taquet14}
{Taquet}, V., {Charnley}, S.~B., \& {Sipil{\"a}}, O. 2014, \apj, 791, 1

\bibitem[{{Terwisscha van Scheltinga} {et~al.}(2022){Terwisscha van
  Scheltinga}, {Ligterink}, {Bosman}, {Hogerheijde}, \&
  {Linnartz}}]{Terwisscha22}
{Terwisscha van Scheltinga}, J., {Ligterink}, N.~F.~W., {Bosman}, A.~D.,
  {Hogerheijde}, M.~R., \& {Linnartz}, H. 2022, \aap, 666, A35

\bibitem[{{Thi} {et~al.}(2010){Thi}, {Woitke}, \& {Kamp}}]{Thi10}
{Thi}, W.-F., {Woitke}, P., \& {Kamp}, I. 2010, \mnras, 407, 232

\bibitem[{{Tielens} \& {Hagen}(1982)}]{Tielens82}
{Tielens}, A.~G.~G.~M. \& {Hagen}, W. 1982, \aap, 114, 245

\bibitem[{{Tielens} {et~al.}(1991){Tielens}, {Tokunaga}, {Geballe}, \&
  {Baas}}]{Tielens91}
{Tielens}, A.~G.~G.~M., {Tokunaga}, A.~T., {Geballe}, T.~R., \& {Baas}, F.
  1991, \apj, 381, 181

\bibitem[{{Tomoda} \& {Kimura}(1983)}]{Tomoda83}
{Tomoda}, S. \& {Kimura}, K. 1983, Chemical Physics Letters, 102, 560

\bibitem[{{Vallet} \& {Masella}(2015)}]{Vallet15}
{Vallet}, V. \& {Masella}, M. 2015, Chemical Physics Letters, 618, 168

\bibitem[{{van Hemert} {et~al.}(2015){van Hemert}, {Takahashi}, \& {van
  Dishoeck}}]{vanHemert15}
{van Hemert}, M.~C., {Takahashi}, J., \& {van Dishoeck}, E.~F. 2015, JPCA, 119,
  6354

\bibitem[{{Vasyunin} {et~al.}(2017){Vasyunin}, {Caselli}, {Dulieu}, \&
  {Jim{\'e}nez-Serra}}]{Vasyunin17}
{Vasyunin}, A.~I., {Caselli}, P., {Dulieu}, F., \& {Jim{\'e}nez-Serra}, I.
  2017, \apj, 842, 33

\bibitem[{{Vasyunin} \& {Herbst}(2013a)}]{Vasyunin13}
{Vasyunin}, A.~I. \& {Herbst}, E. 2013a, \apj, 762, 86

\bibitem[{{Vidali} {et~al.}(1991){Vidali}, {Ihm}, {Kim}, \& {Cole}}]{Vidali91}
{Vidali}, G., {Ihm}, G., {Kim}, H.-Y., \& {Cole}, M.~W. 1991, Surface Science
  Reports, 12, 135

\bibitem[{{Wakelam} {et~al.}(2017){Wakelam}, {Loison}, {Mereau}, \&
  {Ruaud}}]{Wakelam17}
{Wakelam}, V., {Loison}, J.~C., {Mereau}, R., \& {Ruaud}, M. 2017, Molecular
  Astrophysics, 6, 22

\bibitem[{{Walrafen}(2004)}]{Walrafen04}
{Walrafen}, G.~E. 2004, in American Institute of Physics Conference Series,
  Vol. 716, Portable Synchrotron Light Sources and Advanced Applications:
  International Symposium on Portable Synchrotron Light Sources and Advanced
  Applications, ed. H.~{Yamada}, N.~{Mochizuki-Oda}, \& M.~{Sasaki}, 45--48

\bibitem[{{Whittet} {et~al.}(2011){Whittet}, {Cook}, {Herbst}, {Chiar}, \&
  {Shenoy}}]{Whittet11}
{Whittet}, D.~C.~B., {Cook}, A.~M., {Herbst}, E., {Chiar}, J.~E., \& {Shenoy},
  S.~S. 2011, \apj, 742, 28

\bibitem[{{Whittet} {et~al.}(2001){Whittet}, {Gerakines}, {Hough}, \&
  {Shenoy}}]{Whittet01}
{Whittet}, D.~C.~B., {Gerakines}, P.~A., {Hough}, J.~H., \& {Shenoy}, S.~S.
  2001, \apj, 547, 872

\bibitem[{{Whittet} {et~al.}(2010){Whittet}, {Goldsmith}, \&
  {Pineda}}]{Whittet10}
{Whittet}, D.~C.~B., {Goldsmith}, P.~F., \& {Pineda}, J.~L. 2010, \apj, 720,
  259

\bibitem[{{Whittet} {et~al.}(2007){Whittet}, {Shenoy}, {Bergin}, {Chiar},
  {Gerakines}, {Gibb}, {Melnick}, \& {Neufeld}}]{Whittet07}
{Whittet}, D.~C.~B., {Shenoy}, S.~S., {Bergin}, E.~A., {et~al.} 2007, \apj,
  655, 332

\bibitem[{{Willacy} {et~al.}(1994){Willacy}, {Williams}, \&
  {Duley}}]{Willacy94}
{Willacy}, K., {Williams}, D.~A., \& {Duley}, W.~W. 1994, \mnras, 267, 949

\bibitem[{{Williams} {et~al.}(1992){Williams}, {Hartquist}, \&
  {Whittet}}]{Williams92}
{Williams}, D.~A., {Hartquist}, T.~W., \& {Whittet}, D.~C.~B. 1992, \mnras,
  258, 599

\bibitem[{{Yeo} \& {Ford}(1994)}]{Yeo94}
{Yeo}, G.~A. \& {Ford}, T.~A. 1994, Spectrochimica Acta Part A: Molecular
  Spectroscopy, 50, 5

\bibitem[{{Zhao} {et~al.}(2018){Zhao}, {Caselli}, \& {Li}}]{Zhao18}
{Zhao}, B., {Caselli}, P., \& {Li}, Z.-Y. 2018, \mnras, 478, 2723

\bibitem[{{Zuo} {et~al.}(2021){Zuo}, {Li}, \& {Zhao}}]{Zuo21}
{Zuo}, W., {Li}, A., \& {Zhao}, G. 2021, \apjs, 252, 22

\end{thebibliography}

\end{document}